\pdfoutput=1

\documentclass[letterpaper]{revtex4}

\usepackage{graphics}
\usepackage{geometry}                
\usepackage{graphicx}
\usepackage{amssymb}
\usepackage{epstopdf}
\usepackage{amsmath}
\usepackage{alltt}
\usepackage{wrapfig}
\usepackage{multirow}
\usepackage{rotating}

\begin{document}

\title{Invited Review Article: Physics and Monte Carlo Techniques as Relevant to Cryogenic, Phonon and Ionization Readout of CDMS Radiation-Detectors}

\author{Steven W. Leman}
\email{swleman@mit.edu}
\affiliation{Kavli Institute for Astrophysics and Space Research, Massachusetts Institute of Technology, Cambridge, MA 02139, USA}


\begin{abstract}
This review discusses detector physics and Monte Carlo techniques for cryogenic, radiation detectors that utilize combined phonon and ionization readout. A general review of cryogenic phonon and charge transport is provided along with specific details of the Cryogenic Dark Matter Search detector instrumentation. In particular this review covers quasidiffusive phonon transport, which includes phonon focusing, anharmonic decay and isotope scattering. The interaction of phonons in the detector surface is discussed along with the downconversion of phonons in superconducting films. The charge transport physics include a mass tensor which results from the crystal band structure and is modeled with a Herring Vogt transformation. Charge scattering processes involve the creation of Neganov-Luke phonons. Transition-edge-sensor (TES) simulations include a full electric circuit description and all thermal processes including Joule heating, cooling to the substrate and thermal diffusion within the TES, the latter of which is necessary to model normal-superconducting phase separation. Relevant numerical constants are provided for these physical processes in germanium, silicon, aluminum and tungsten. Random number sampling methods including inverse cumulative distribution function (CDF) and rejection techniques are reviewed. To improve the efficiency of charge transport modeling, an additional second order inverse CDF method is developed here along with an efficient barycentric coordinate sampling method of electric fields. Results are provided in a manner that is convenient for use in Monte Carlo and references are provided for validation of these models.
\end{abstract}

\maketitle
\tableofcontents

\section{Introduction}

Cryogenic radiation-detectors that utilize ionization, phonon and / or scintillation measurements are being used in a number of experiments. Both the Cryogenic Dark Matter Search (CDMS)~\cite{Ahmed2010, Ahmed2011} and EDELWEISS~\cite{Armengaud2010} dark matter search utilize silicon and / or germanium targets to detect recoils of radiation in the target masses. A combination of ionization and phonon readout is used to provide discrimination of gamma- and neutron-recoil types. The CRESST dark matter search utilizes CaWO$_4$ targets and readout scintillation and phonon signal to discriminate between recoil types. The advantage of reading out both phonon and ionization (or scintillation) signals comes about from the differing ratios of ionization and phonon energy or scintillation and phonon energy created in electron- and nuclear-recoils in the detectors. The ratio of these two energies leads to a powerful discriminator for the experiment's desired recoil type.

Both the ionization and phonon readout can be used to generate position estimators for the initial radiation interaction, leading to fiducial volume selection. In the ionization signal this is generally accomplished by instrumenting different parts of the detector with independent readout channels and vetoing events with large signal contribution outside of the desired fiducial volume. In the phonon signal it is generally required to measure the early, athermal component of the phonon signal which still retains a position dependent component.

The physics required to accurately model these detectors is presented in this paper along with appropriate numerical tricks that are useful for an efficient detector Monte Carlo. This paper proceeds with a review of radiation interactions, charge transport physics, phonon transport physics, instrumentation. Monte Carlo techniques and relevant physical constants are included where appropriate. 

This paper will focus on the use of silicon and germanium detector masses, both of which are group IV semiconductors. However there are other relevant materials in use such as calcium tungstate (CaWO$_4$) which leads to a small loss of generality.

\subsection{The CDMS Experiment and Detectors}

The Cryogenic Dark Matter Search~\cite{Ahmed2010, Ahmed2011} utilizes silicon and germanium detectors to search for Weakly Interacting Massive Particle (WIMP) dark matter~\cite{Spergel2007,Tegmark2004} candidates. The silicon or germanium nuclei provide a target mass for WIMP-nucleon interactions. Simultaneous measurement of both phonon energy and ionization energy provide a powerful discriminator between electron-recoil interactions and nuclear-recoil interactions. Background radiation primarily interacts through electron-recoils whereas a WIMP signal would interact through nuclear-recoils. The experiment is located in the Soudan Mine, MN, U.S.A.

\begin{figure}[h]
\begin{tabular}{c}
\begin{minipage}{0.5\hsize}
\begin{center}
\includegraphics[width=7cm]{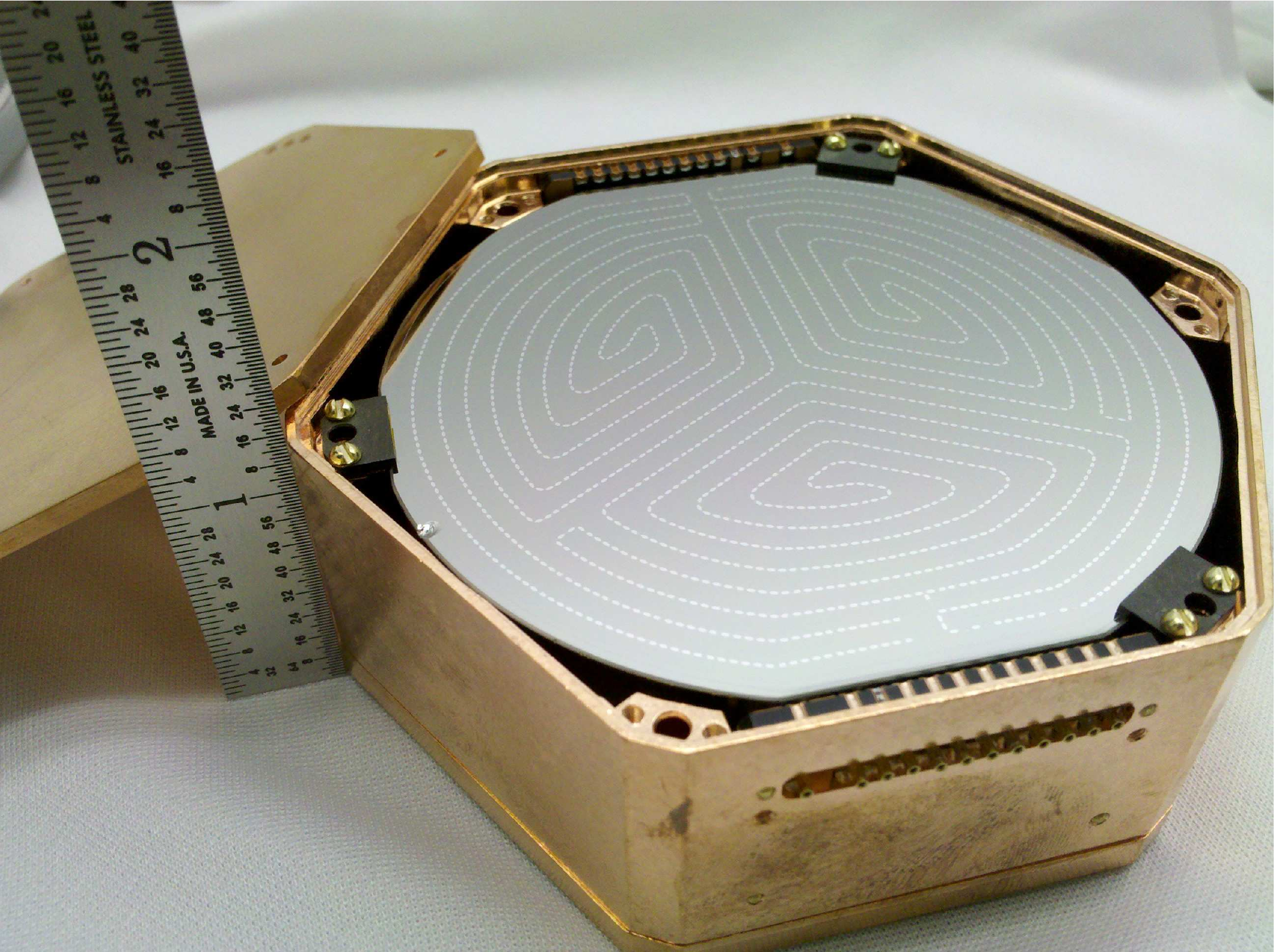}
\end{center}
\end{minipage}
\begin{minipage}{0.45\hsize}
\begin{center}
\includegraphics[width=7cm]{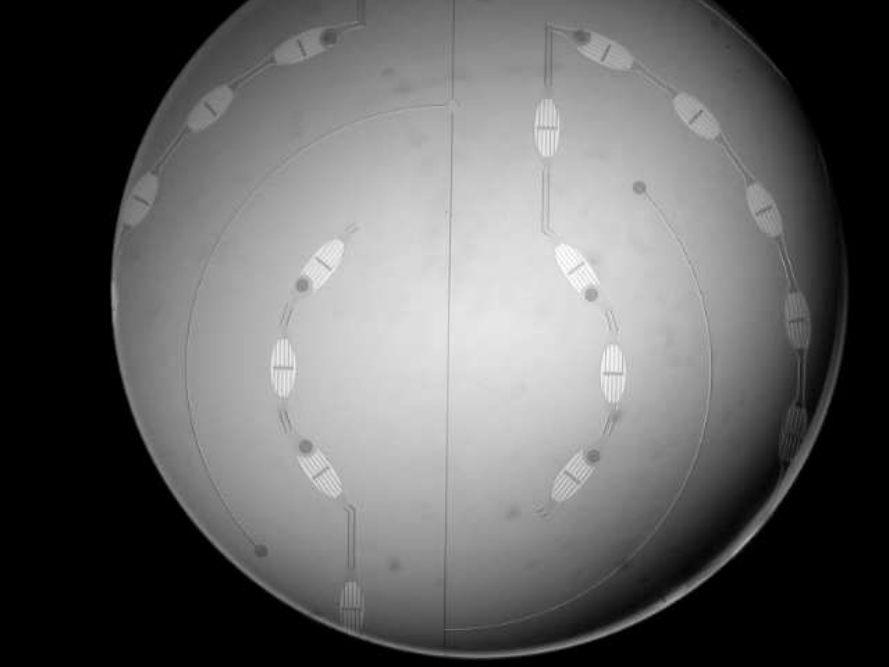}
\end{center}
\end{minipage}
\end{tabular}
\caption[] { \label{fig:iZip}(left) A CDMS ``iZIP'' detector with photolithographically defined phonon sensors. The crystal is 3~inches in diameter and mounted in its copper housing. The top surface contains an outer, guard phonon sensor and three inner phonon sensors from which an event's position estimate can be made. The opposite face (not shown) has a similar channel design, but rotated 60~degrees.}
\caption[] { \label{fig:iZipTES}(right) Close-up view of the iZIP phonon channel and ionization channel (thin lines in between the phonon sensors). The phonon channel is held at ground and the ionization channel is held at $\sim \pm$2~V for the top (bottom) surfaces.}
\end{figure}

The most recent phase of the CDMS experiment has involved fabrication, testing and commissioning of large, 3~inch diameter, 1~inch thick [100] germanium crystals. The CDMS-iZIP (interleaved Z--dependent Ionization and Phonon) detectors are 3~inches in diameter and 1~inch thick with a total mass of about 607~grams~\cite{Brink2006}. The iZIP detector utilizes both anode and cathode lines on the same side of the detector similar to a Micro-Strip Gas Chamber (MSGC)~\cite{Knoll, Oed1988, Luke1994, Brink2006} as shown in Figure~\ref{fig:iZip} and~\ref{fig:iZipTES}. Unlike an MSGC however, there is a set of anode and cathode lines on both sides of the detector. This ionization channel design is used to veto events interacting near the detector surfaces. An amorphous silicon layer, deposited under the metal layers, increases the breakdown voltage of the detectors. The total iZIP aluminum coverage is $\sim$4.8\% active and $\sim$1.5\% passive per side.

\section{Radiation Modeling}

When using a Monte Carlo of a detector, it is often helpful or necessary to have a numerical model of radiation interactions in the detector. Many readers will find it valuable to use separate modeling software such as GEANT4~\cite{Agostinelli2003}. A brief description of these interactions follows.

Low energy gamma-rays (x-rays) predominantly interact via photoelectric absorption in which all of the gamma-ray energy is deposited in a single interaction location. High energy gamma-rays interact via Compton scattering in which some of the gamma-ray's initial energy is transferred to an electron and the gamma-ray continues along a different trajectory with reduced energy. The gamma-ray will generally continue to scatter until it leaves the detector volume or terminates with a photoelectric absorption. In silicon (germanium), for photon energies greater than 60 (160)~keV, Compton scattering dominates~\cite{Knoll, RPP}.

Both of these electron interactions result in a high energy electron being produced which then undergoes a rapid cascade process resulting in a large number of electron-hole pairs~\cite{RPP, Cabrera1993}. This initial cascade process ceases around the scale of the mean electron-hole pair creation energy ($E_{eh,create}$) resulting in an expected number of electron-hole pair $n_{eh} = E_\gamma / E_{eh,create}$. Due to correlations in the cascade process, the variance in the number of electron-hole pairs is reduced, relative to Poisson statistics, and given by $\sigma_{eh}^2 = F \times n_{eh}$, where $F$ is the Fano factor~\cite{Fano1947}. These high energy electron-hole pairs will then shed phonons until they reach the semiconductor gap energy $E_{\text{gap}}$ which results in the fraction $1-E_{\text{gap}} / E_{eh,create}$ of energy in prompt phonons and the remainder in the electron-hole pair system.

\begin{figure}
\begin{center}
\includegraphics[width=14cm]{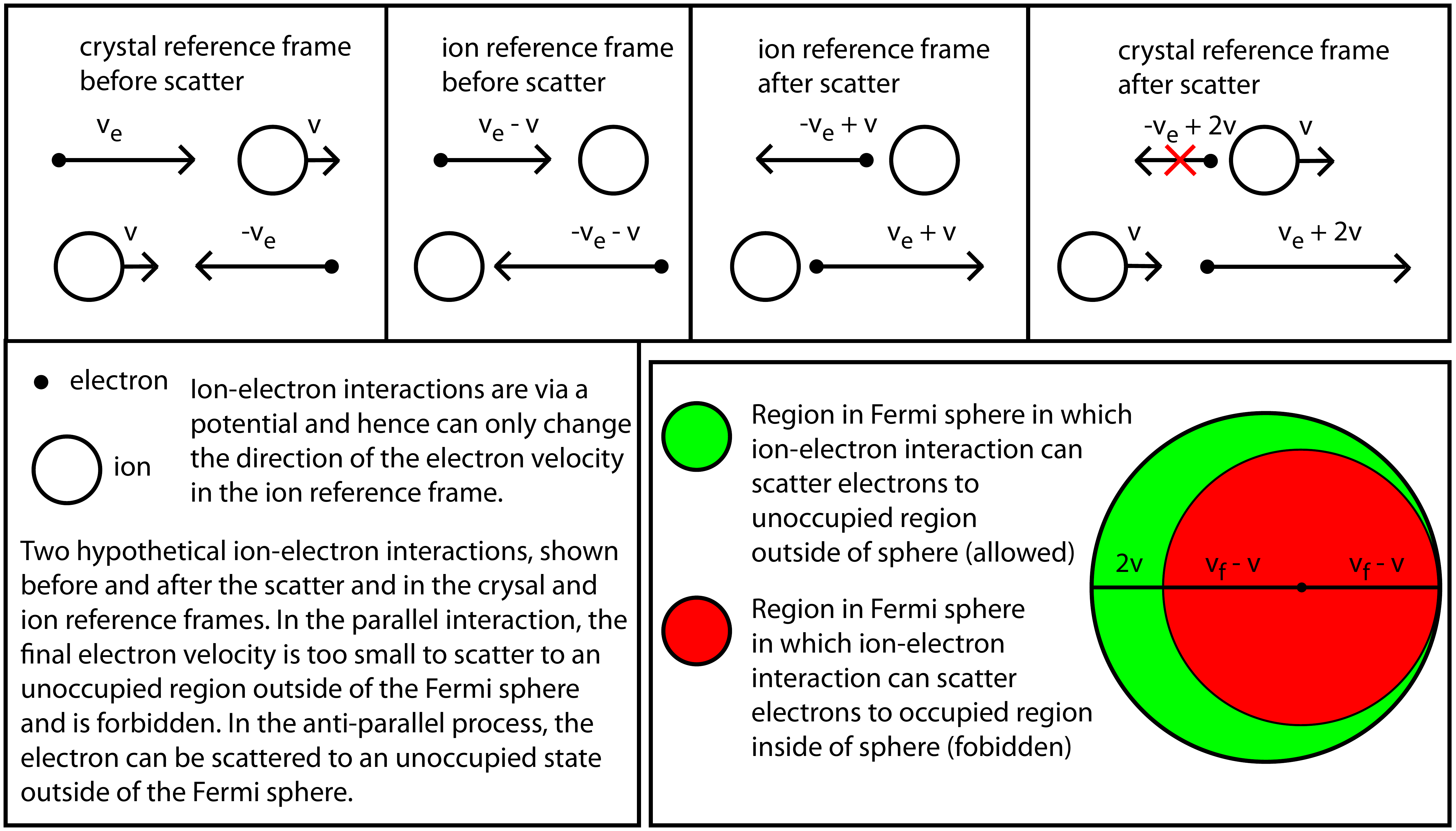}
\end{center}
\caption[] { \label{fig:ElecIonScatter}
Description of ion-electron potential interaction process which shows that for ion velocity much less than the Fermi velocity ($v \ll v_f$) the number of electron states that an ion can interact with scales like $v$.}
\end{figure}

Neutrons that interact in the detector bulk will knock an ion out of its lattice site and displace it to some other location in the crystal. This high energy ion will interact with both the lattice ions and / or valence electrons, with competing cross sections, before reaching some other location in the crystal. Interactions with the lattice ions can be described to first approximation as Rutherford scattering~\cite{Rutherford1911} with differential energy loss per unit length $-dE/dx \sim v^{-2}$, where $v$ is the ion's velocity~\cite{Bonderup1978}. For interactions with valence electrons, the number of electron states which can be excited to an accessible state outside of the Fermi sphere scales like velocity $v$ hence the differential energy loss per unit length scales like $-dE/dx \sim v$~\cite{Bonderup1978}. A description of this process is shown in Figure~\ref{fig:ElecIonScatter}. There are important screening and velocity dependent cross sections in both energy loss lengths~\cite{Bonderup1978, Lindhard1954, Lindhard1963, Lindhard1963_2, Jones1975} but to first approximation these equations show the velocity dependent competition in the two scattering rates. Whichever interaction occurs first, there is generally a cascade of many scattering events resulting in a larger amount of energy being deposited in the ion lattice compared to gamma-ray interactions. For historical reasons, the reduced amount of energy in the electron-hole (ionization) system, compared to an equal energy deposition by a gamma-ray,  has brought about the term \emph{nuclear quenching} to describe the reduced ionization signal. The details of the cascade and reduction in the number of electron-hole pairs is described by Lindhard theory~\cite{Lindhard1963} and given as $f = k g(\epsilon) / (1 + k g(\epsilon))$ where $\epsilon = 11.5 E_R Z^{-7/3}$, $k=0.133 Z^{2/3} A^{-1/2}$ and $g(\epsilon) = 3  \epsilon^{0.15} + 0.7 \epsilon^{0.6} + \epsilon$~\cite{Lewin1996}. The recoil energy is $E_R$ and given in units of keV, $Z$ is the atomic number and $A$ is the atomic mass.

Beta radiation represents another class of electron-recoil interactions. The attenuation lengths are much shorter, however, resulting in a class of events sometimes referred to as \emph{surface interactions}. These surface interactions can result in signals that differ from bulk events of the same energy and electron- / nuclear-recoil type. For example, the events may be located in regions of large electric fringing fields, directing charges away from readout electrodes or near mounting hardware that absorbs scintillation light.

\section{Phonon Simulation}

\subsection[Phonon Introduction]{Introduction} 
Phonons, in the context of this review, are quantized vibration modes that exist in periodic structures such as silicon and germanium crystals. They are excitations which, in the lattice, and in materials sufficiently cooled that charge carriers are frozen out, mediate thermal transport. They are described by the Hamiltonian

\begin{equation}
H = \sum_{\substack{i}}{\frac{p_i^2}{2 m}} + \sum_{\substack{i,j}}{\frac{m \omega ^2}{2} (x_i-x_j)^2},
\end{equation}
where $m$ is the mass of the lattice atoms and $\omega$ is the frequency of oscillations about the center of the harmonic potential between an atom and its nearest neighbor atom~\cite{Kittel, Ashcroft, Wolfe}.

In one dimension, the solution to the time-independent Schrodinger equation $ H \left| x \right> = E  \left| x \right> $  yields the different oscillation modes at atom $j$

\begin{equation}
x_j \sim e^{i k_n j a}
\end{equation}
where $i = \sqrt{-1}$ here, the wave number $k_n = \frac{2 n \pi}{N a}$, $a$ is the lattice spacing and $N$ are the number of lattice sites.

In Monte Carlo, phonons are simply treated as non interacting particles with decay and mass defect scattering properties described in the remainder of this chapter. In general they have a nonlinear dispersion relationship; however, due to rapid down conversion to lower frequencies they spend most of their time in an energy region with linear dispersion relationship. Hence, a linear dispersion relation is sufficient for phonon transport modeling in these detectors.

\subsection{Prompt Phonon Distributions} 

Immediately after the recoil, a population of prompt phonons exist in the immediate vicinity of the interaction point. Particular details of the frequency distribution and mode population are not well known but we can make a few deductions, explained in more detail in the following sections, that will lead us an initial distribution for Monte Carlo.

Anharmonic decay, due to nonlinear terms in the elastic coupling between adjacent lattice ions, causes the phonons to rapidly down convert into a lower frequency distribution. This process allows us to start a Monte Carlo with any high frequency distribution and details of the distribution will rapidly be lost; we use the Debye frequency as a naive starting point. Isotope scattering, which also occurs at a high rate for a high frequency phonons, causes the phonons to obtain their equilibrium mode density; we use the equilibrium mode density as a naive starting point. The approximations are good in the sense that the detector's phonon response is insensitive to variations in the distributions.

They are valid since the detectors are large compared to the initial characteristic interaction lengths of phonons. Furthermore, later generations of phonons are ballistic and timing information in the measured phonon pulses is determined by the detector geometry and the loss rate of phonons at surfaces.

\subsection[Phase Velocities]{Phase Velocities and Polarization Vectors}

The so called phase velocity surfaces represent the direction dependent phonon phase velocity $\vec{v}_p = \vec{v}_p (\theta, \phi)$ . In general they are given by the eigenvalue Equation~\ref{eq:PhaseVelocity}.

\begin{equation}\label{eq:PhaseVelocity}
\rho \omega^2 \epsilon_\mu = \sum_{\tau} \left(\sum_{\sigma \nu}
c_{\mu \sigma \nu \tau} k_\sigma k_\nu \right) \epsilon_\tau,
\end{equation}
where $\rho$ is the crystal's mass density,\\
$\omega$ is the phonon frequency,\\
$\epsilon_\mu$ is a component of the polarization vector
$\mathbf{\epsilon}$,\\
$c_{\mu \sigma \nu \tau}$ are components of the elastic constant tensor and,\\
$k_{\sigma}$ is a component of the phase velocity vector
$\mathbf{k}$~\cite{Kittel}.

Not all of the elastic constants are independent, reducing via a Voigt contraction~\cite{Nye} the number that we need to keep track off. Additionally, symmetries in a cubic crystal allow for further reduction in components and we can define three independent constants $C_{11} = c_{xxxx}= c_{yyyy}= c_{zzzz}$, $C_{12} = c_{xxyy}= c_{yyzz}= c_{zzxx}$ and $C_{44} = c_{xyxy}= c_{yzyz}= c_{zxzx}$. This contraction simplifies Equation~\ref{eq:PhaseVelocity} significantly~\cite{Ashcroft} and we are left solving matrix~\ref{eq:PhaseVelocityMatrix} for its eigenvectors and eigenvalues.

\begin{align}\label{eq:PhaseVelocityMatrix}
\left(
  \begin{array}{ccc}
    C_{11} k_x k_x + C_{44} (k_y k_y + k_z k_z) & (C_{12} + C_{44}) k_x k_y &   \\
    (C_{12} + C_{44}) k_x k_y & C_{44} (k_x k_x + k_z k_z) + C_{11} k_y k_y &   \cdots \hspace{20 mm}\\
    (C_{12} + C_{44}) k_x k_z & (C_{12} + C_{44}) k_y k_z & \\
  \end{array}
\right.
\notag\\
\left.
  \begin{array}{ccc}
  & & (C_{12} + C_{44}) k_x k_z \\
  & & (C_{12} + C_{44}) k_y k_z \\
  & & C_{44} (k_x k_x + k_y k_y) + C_{11} k_z k_z \\
  \end{array}
\right)
\end{align}

\begin{figure}
\begin{center}
\includegraphics[width=14cm, bb=0 0 1200 900]{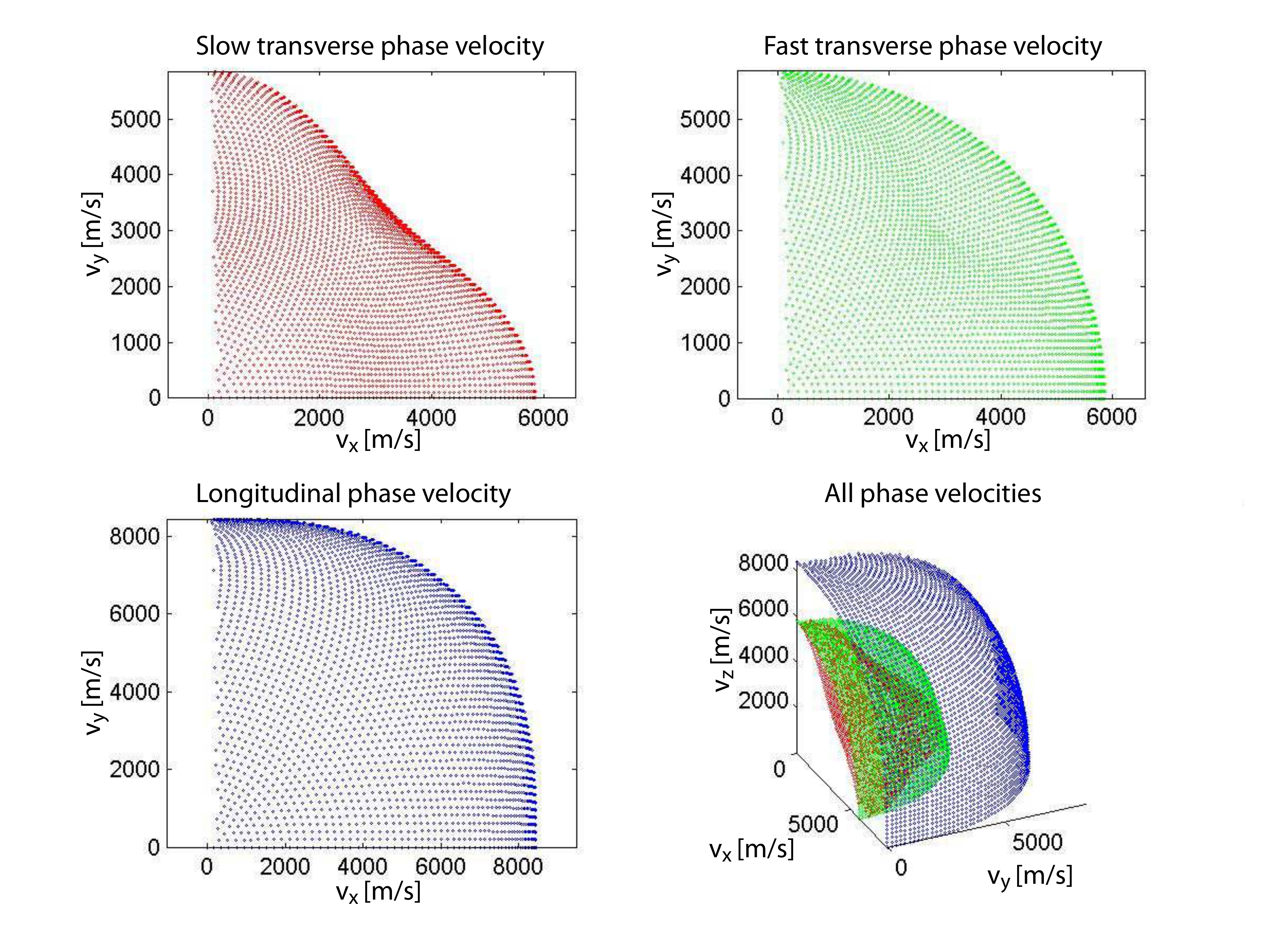}
\end{center}
\caption[Phase Velocities in Silicon] { \label{fig:SiDispersion}
Phase velocities in silicon. The distance from the origin represents the phase velocity speed (in m/s). The first three plots are views from the north pole. The fourth view containing all three surfaces is offset from the north pole. In the plots all modes are equally populated, which does not reflect the actual mode populations.}
\end{figure}

The eigenvectors represent the three polarization vector directions and the three eigenvalues equal $\rho \omega^2$ for the longitudinal, slow-transverse and fast-transverse modes. The anisotropy in the cubic silicon crystal leads to the phase velocity surfaces being non-spherical (see Figure~\ref{fig:SiDispersion}). The phase velocities are used to determine both the group velocities (Section~\ref{sect:GroupVelocity}) and also the isotope scattering rates.

\subsection{Group Velocities}
\label{sect:GroupVelocity} Phonon group velocities are found by
solving
\begin{equation} \label{eq:GroupVelocity}
\vec{v}_g (\theta, \phi) = \frac{\partial \omega ( \theta, \phi ) }
{\partial \vec{k}}.
\end{equation}

\begin{figure}
\begin{center}
\includegraphics[width=14cm, bb=0 0 1200 900]{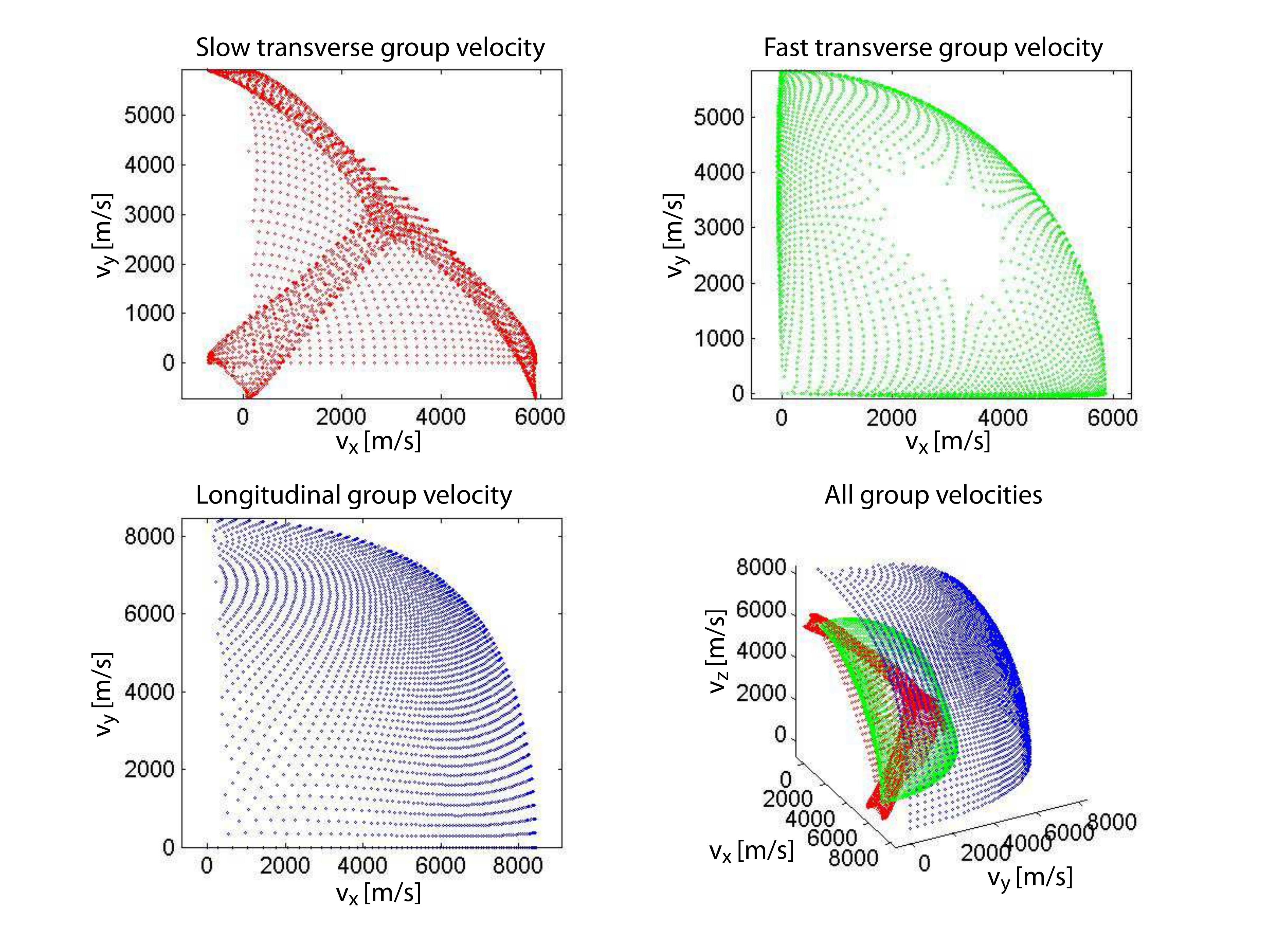}
\end{center}
\caption[Group Velocities in Silicon] { \label{fig:SiGroupVelocity}
Group velocities in silicon. Energy is focused in the direction of heavy banding and leads to the term \emph{phonon focusing}. The distance from the origin represents the speed that phonon energy is carried through the crystal (in m/s). In the plots all modes are equally populated, which does not reflect the actual mode populations.}
\end{figure}

The slight lack of sphericity in the phase velocity surfaces (see Figure~\ref{fig:SiDispersion}) has a very dramatic effect on the transverse phonon group velocities~\cite{Northrop1980, Tamura1991, Maris1993, Tamura1993LT} (see Figure~\ref{fig:SiGroupVelocity}). The longitudinal phonon's group velocity is only mildly affected. Energy is focused in the direction of heavy banding and leads to the term \emph{phonon focusing}. The point density in the plots is misleading as the three modes are shown to be equally populated. Isotope scattering including anisotropic scattering rates  (see Section~\ref{sect:IsotopeScatter}) leads to the phonon modes in silicon being populated as follows: slow-transverse (55\%), fast-transverse (35\%) and longitudinal (10\%).

\subsection{Anisotropic Isotope Scattering} \label{sect:IsotopeScatter}
\begin{figure}
\begin{center}
\includegraphics[width=10cm, bb=0 0 1182 508]{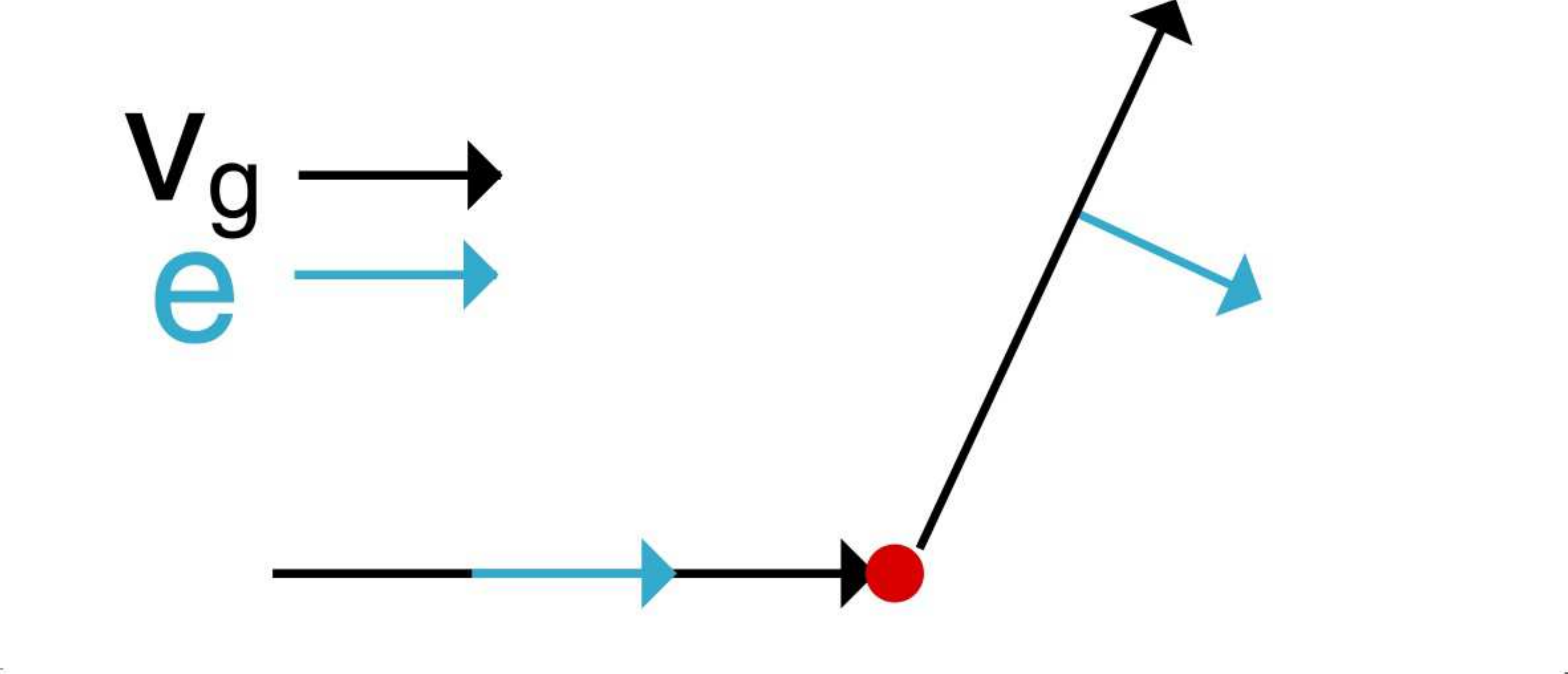}
\end{center}
\caption[Anisotropic Isotope Scattering] {
\label{fig:isotopeScatter} Phonons isotope scatter off mass defects in the crystal. Equation~\ref{eq:MicroIsotopeScatter} gives the individual phonon scatter rates. $\vec{v_g}$ is the group velocity and $\vec{e}$ is the polarization vector.}
\end{figure}

Phonons scatter off mass defects in the crystal (see Figure~\ref{fig:isotopeScatter}). Additionally, they can change modes. The bulk scattering rate is given by
\begin{equation}
\Gamma_I = B \nu^4[s^3],
\end{equation}
where $\nu$ is the phonon frequency and $B$ is a scattering rate constant~\cite{Tamura1991, Tamura1993LT, Maris1990, Msall1993, Tamura1993} (see Table~\ref{tab:phononSimConst}). The scattering rate for individual phonons is given by
\begin{equation}
\label{eq:MicroIsotopeScatter}\gamma \sim \frac{ |\vec{e}_{\lambda}
\cdot \vec{e}_{\lambda ^\prime}|^2} {\nu^3_{\lambda ^\prime}},
\end{equation}
where $\nu_{\lambda ^\prime}$ is the final state phonon frequency in Hz, $\vec{e}$ is the polarization vector, $\lambda$ represents the initial phonon and $\lambda ^\prime$ represents the outgoing phonon~\cite{Tamura1993LT, Tamura1993}. It is the dot product in Equation~\ref{eq:MicroIsotopeScatter} which allows mode mixing and the denominator which ensures the correct populations in the ratios. In silicon the populations when including anisotropic scattering rates are slow-transverse (55\%), fast-transverse (35\%) and longitudinal (10\%). The standard treatment is to determine if a phonon isotope scatters via $\Gamma_I$ and then determine its polarization and direction via $\gamma$. After the initial anharmonic decay has settled down, isotope scattering dominates.

This process is unfortunately computationally expensive due to sample-rejection techniques~\cite{NumericalRecipes} and after several iterations an isotropic scattering process can be used for individual phonons with little loss of modeling accuracy.

\subsection{Anharmonic Decay}
\label{sect:AnharmonicDecay}

\subsubsection{General Case}
\begin{figure}
\begin{center}
\includegraphics[width=10cm, bb=0 0 557 380]{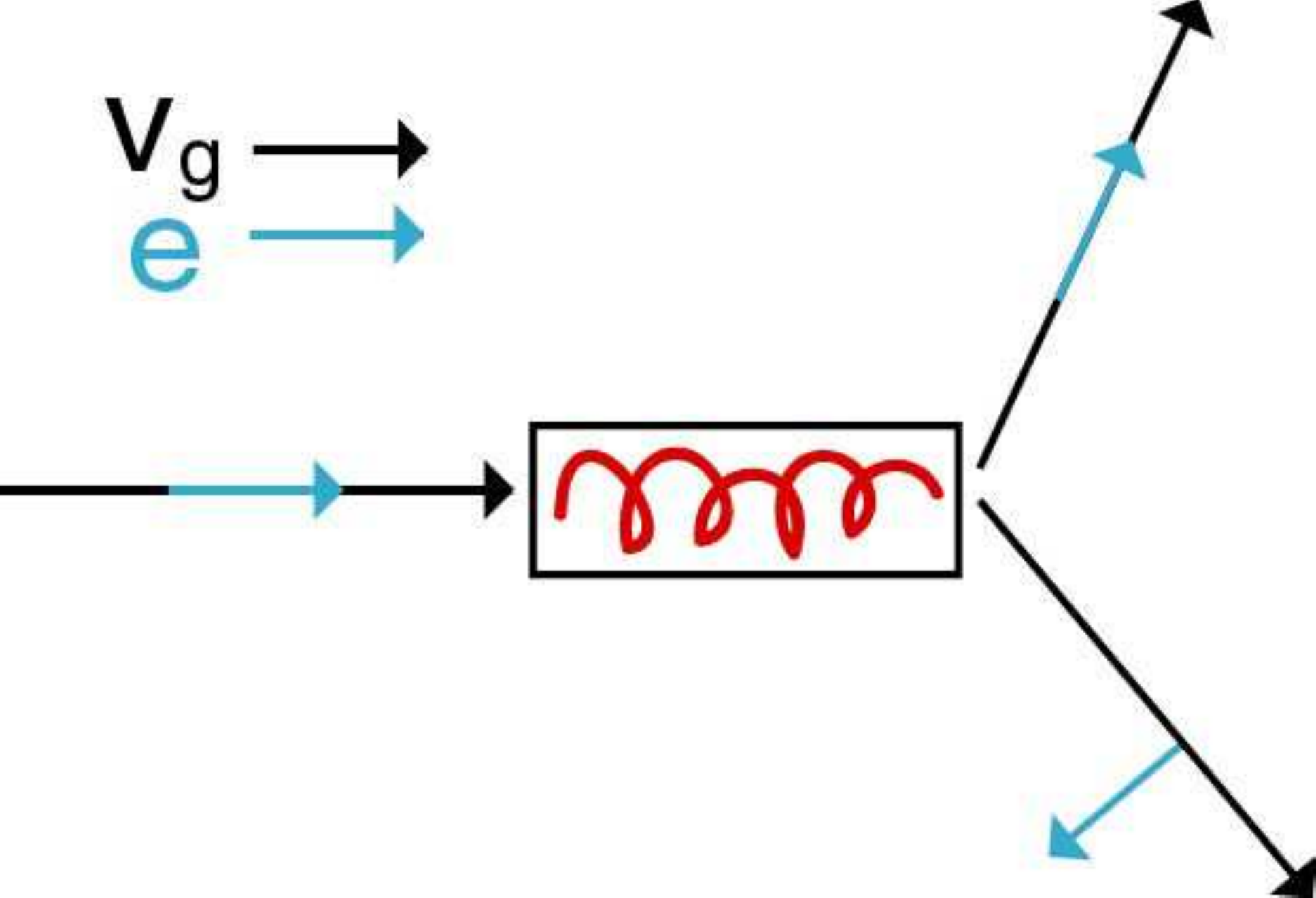}
\end{center}
\caption[Anharmonic Decay] { \label{fig:anharmonicDecay}
Longitudinal phonons decay due to nonlinear terms in the elastic coupling constants. $\vec{v_g}$ is the group velocity and $\vec{e}$ is the polarization vector.}
\end{figure}

Nonlinear terms in the elastic coupling constants cause a longitudinal phonon to down convert to two lower energy phonons (see Figure~\ref{fig:anharmonicDecay}). The bulk decay rate is given by
\begin{equation}
\Gamma_A = A \nu^5 [s^4],
\end{equation}
where $\nu$ is the phonon frequency and $A$ is a decay rate constant~\cite{Tamura1993LT, Maris1990, Msall1993, Tamura1993} (see Table~\ref{tab:phononSimConst}). The decay rate for transverse phonons is negligible~\cite{Tamura1985TA}. The three body problem requires that both energy and momentum are conserved. These conditions make an exact solution computationally prohibitive for large amounts of phonons.

\subsubsection{Isotropic Approximation}
To allow computations to proceed in a finite time, an exact solution to anharmonic decay is abandoned and an isotropic approximation is used (the full anisotropic phase velocities and group velocities are still easily used for isotope scattering and phonon transport). The energy distribution calculations are still difficult but fortunately have already been carried out~\cite{Tamura1985, CabreraTamura}. Once the energies have been determined, calculating the resultant scattering angles based on energy and momentum conservation is fairly straightforward. Due to the different energy-momentum dispersion relations for the longitudinal and transverse phonons, there are two different decay branches, $L\rightarrow L^\prime+T^\prime$ and $L\rightarrow T^\prime_1+T^\prime_2$. The $L\rightarrow L^\prime+T^\prime$ energy distribution is given by
\begin{equation}
\Gamma_{L\rightarrow L^\prime+T^\prime} \sim
\frac{1}{x^2}\left(1-x^2\right)^2 \left[(1+x)^2 - \delta^2
(1-x)^2\right] \left[1 + x^2 - \delta^2 (1-x)^2\right]^2,
\end{equation}
where $x = E_{L^\prime} / E_L$,\\
$\delta = \frac{v_l}{v_t}$ and,\\
$\frac{\delta-1}{\delta+1} < x < 1$.

This approximation results in the outgoing phonons having an angular displacement from the initial phonon given by
\begin{equation}
\cos(\theta_{L^\prime}) = \frac{1+x^2-\delta^2(1-x)^2}{2 x},
\end{equation}
\begin{equation}\label{eq:L2LTAngleT}
\cos(\theta_{T^\prime}) = \frac{1-x^2+\delta^2(1-x)^2}{2 \delta
(1-x)}.
\end{equation}

The $L\rightarrow T^\prime_1+T^\prime_2$ energy distribution is given
\begin{equation}\label{eq:L2TTDecayProb}
\Gamma_{L\rightarrow T^\prime_1+T^\prime_2} \sim \left(A + B \delta
x - B x^2\right)^2 + \left[C x (\delta - x) -
\frac{D}{\delta-x}\left(x-\delta-\frac{1-\delta^2}{4
x}\right)\right]^2,
\end{equation}
where $x = \delta \frac{E_{T^\prime_1}}{E_{L}}$,\\
$\delta = \frac{v_l}{v_t}$,\\
$\frac{\delta-1}{2} < x < \frac{\delta+1}{2}$,\\
$A = \frac{1}{2} (1-\delta^2)[\beta + \lambda + (1+\delta^2)(\gamma + \mu)]$,\\
$B = \beta + \lambda + 2 \delta ^2 (\gamma + \mu)$,\\
$C = \beta + \lambda + 2 (\gamma + \mu)$ and,\\
$D = (1-\delta^2)(2 \beta + 4 \gamma + \lambda + 3 \mu)$.\\
The constants $\gamma$ and $\mu$ are the Lam\'e constants and $\beta$ and $\gamma$ are third order elastic constants in an isotropic model (there is additionally a third independent elastic constant $\alpha$ but it drops out of the equations). 

\begin{figure}
\begin{tabular}{c}
\begin{minipage}{0.95\hsize}
\begin{center}
\includegraphics[width=10cm]{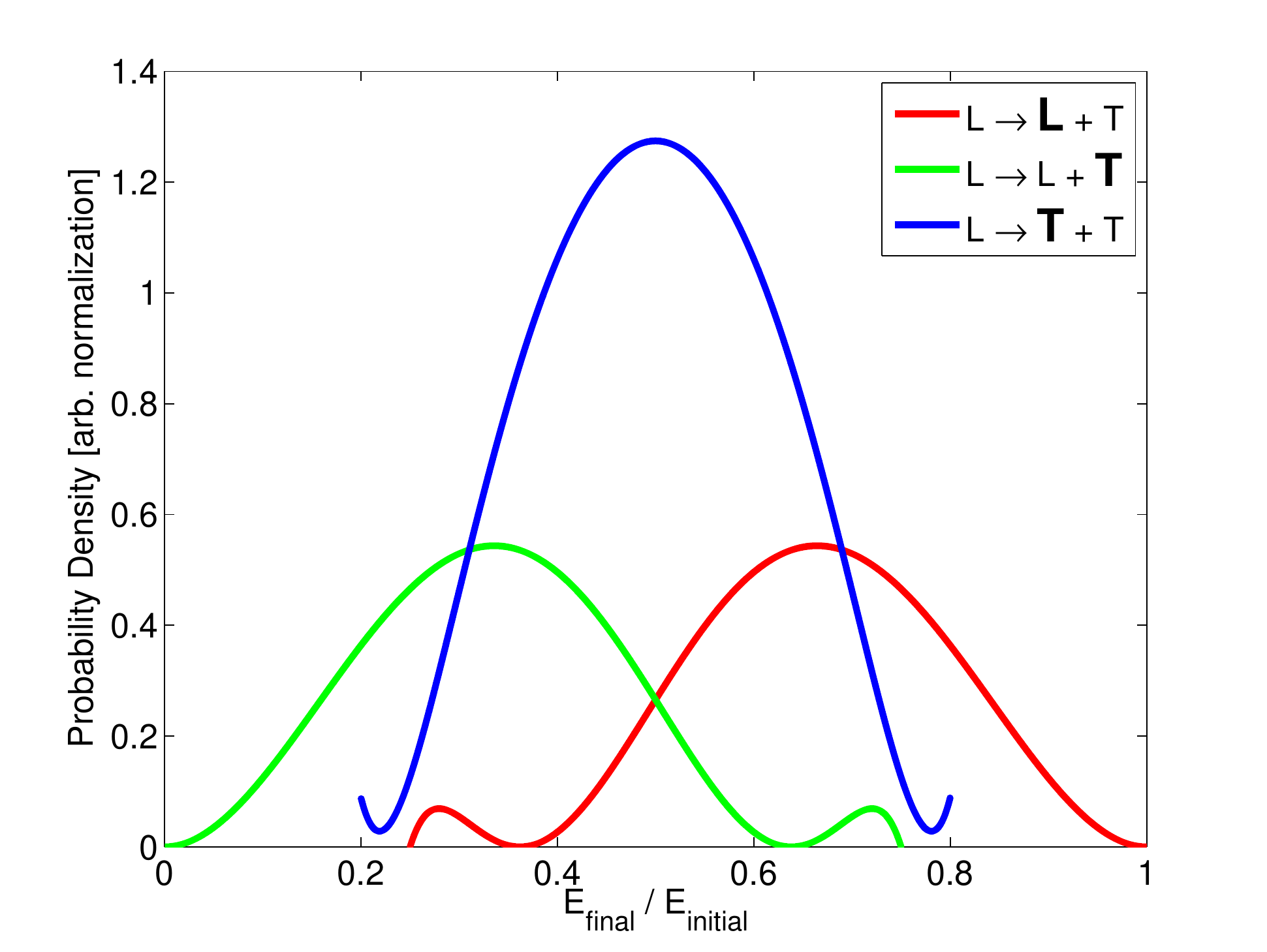}
\end{center}
\caption[Anharmonic Decay Daughter Energies in Germanium] {
\label{fig:anhEnergyGe} Resultant energies in longitudinal phonon decay in germanium. The two branches $L\rightarrow L+T$ and $L\rightarrow T+T$ are shown; the plotted distribution is indicated in bold face in the legend.}
\end{minipage}\\
\\
\begin{minipage}{0.95\hsize}
\begin{center}
\includegraphics[width=10cm]{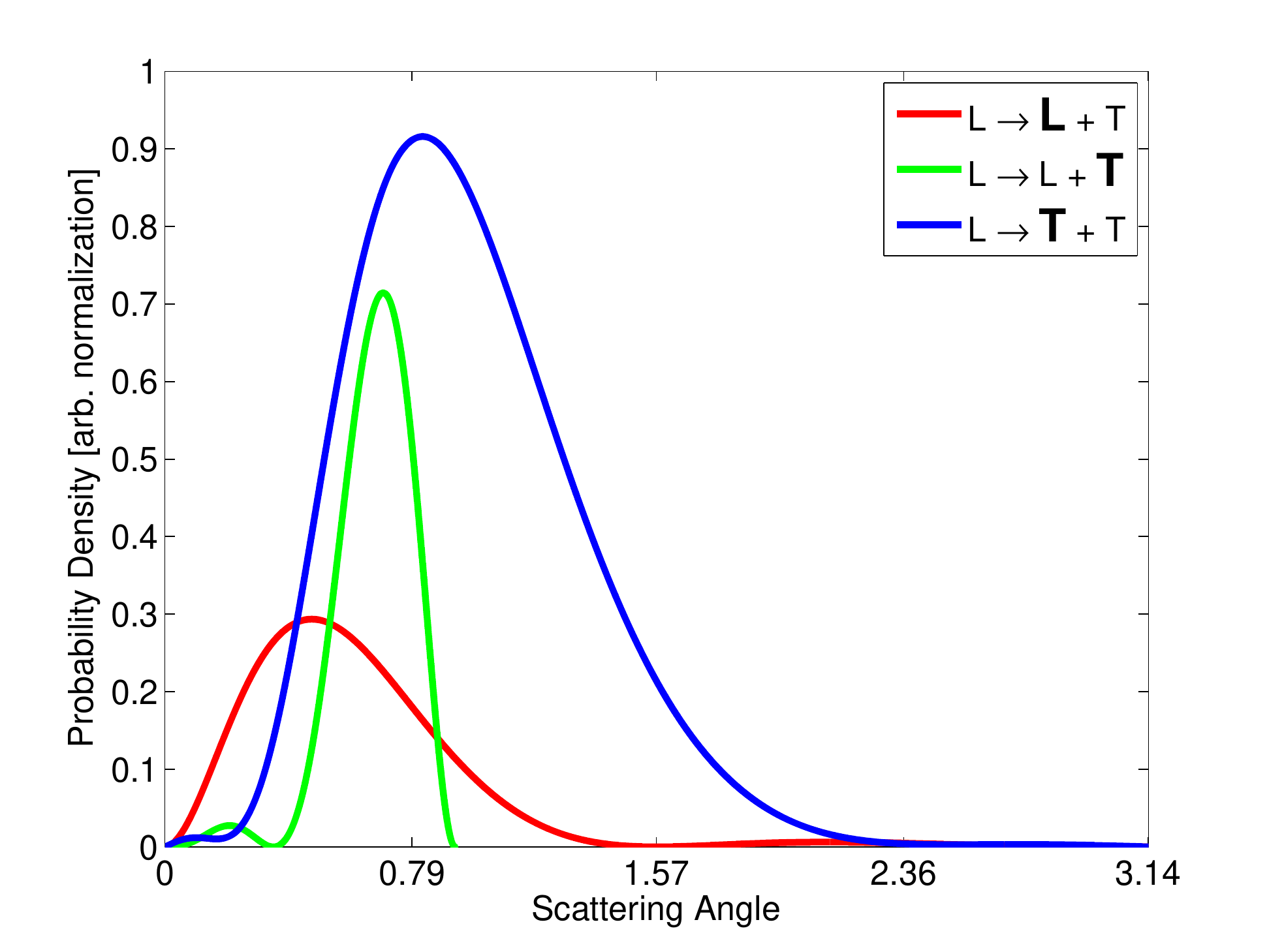}
\end{center}
\caption[Anharmonic Decay Daughter Angles in Germanium] {
\label{fig:anhAngleGe} Resultant angles in longitudinal phonon decay in germanium. The two branches $L\rightarrow L+T$ and $L\rightarrow T+T$ are shown; the plotted distribution is indicated in bold face in the legend.}
\end{minipage}
\end{tabular}
\end{figure}

This approximation results in the outgoing phonons having an angular displacement from the initial phonon given by Equations~\ref{eq:L2TTAngle1} and~\ref{eq:L2TTAngle2}.

\begin{equation}\label{eq:L2TTAngle1}
\cos(\theta_{T^\prime_1}) = \frac{1-\delta^2(1-x)^2+\delta^2 x^2}{2
\delta x},
\end{equation}

\begin{equation}\label{eq:L2TTAngle2}
\cos(\theta_{T^\prime_2}) = \frac{1-\delta^2 x^2+\delta^2 (1-x)^2}{2
\delta (1-x)},
\end{equation}
where the trivial substitution $x \rightarrow 1-x$ is made for the second phonon. With these closed-form energy densities and scattering angles, plots can be generated to aid understanding of these events (see Figures~\ref{fig:anhEnergyGe} and~\ref{fig:anhAngleGe}).

These angles are relative the initial momentum vector $\vec{k}$ and need to be converted into the Monte Carlo coordinate system. Polar coordinates are useful and angles are provided in this system. In addition to rotation angles $\theta_{L^\prime}$ and $\theta_{T^\prime}$  (or $\theta_{T^\prime_1}$ and $\theta_{T^\prime_2}$) an additional azimuth angle, relative to the initial momentum vector $\vec{k}$, and in the isotropic approximation is randomly distributed from $[0, 2\pi]$, is specified as $\theta_{2\pi}$.

In terms of initial elevation and azimuth angles $\Phi$ and $\Theta$, scattering angles $\theta_{T^\prime_1}$ and $\theta_{T^\prime_2}$ and azimuth scattering angle $\theta_{2\pi}$, the final angles $\Phi_{1}$ and $\Theta_{1}$ that describe the phonon momentum vectors $\vec{k}_1$ are 

\begin{equation}
\Phi_{1} = \arccos [-\sin{\Phi} \sin{\theta_{T^\prime_1}} \cos{\theta_{2\pi}} + \cos{\Phi} \cos{\theta_{T^\prime_1}}]
\end{equation}
and
\begin{equation}
\Theta_{1} = \Theta - \text{arctan2} [-\sin{\theta_{T^\prime_1}} \sin{\theta_{2\pi}} \sin{\Phi} , \cos{\theta_{T^\prime_1}} - \cos{\Phi} \cos{\Phi_{1}}].
\end{equation}

The final angles $\Phi_{2}, \Theta_{2}$ for the other phonon momentum vector $\vec{k}_2$ are found by replacing $\theta_{T^\prime_1}$ with $\theta_{T^\prime_2}$ and $\theta_{2\pi}$ with $2\pi - \theta_{2\pi}$.

\subsection{Phonon Losses at Surfaces}
Eventually the phonons will interact at surfaces where they are instrumented, reflected back into the crystal, down converted to a lower energy or are lost to the environment. 

Phonons can reflect either specularly or diffusively from the surfaces. Specular reflection can occur on smooth, untreated semiconductor wafer surfaces. It is the simplest to describe with incident and reflected angles relative to the normal equal, $\theta_i = \theta_r$.

Diffusive scattering is also common on surfaces that have been damaged or roughened during fabrication. In the ideal case the scattering angle is described by Lambert's cosine law where the angular distribution scales like $\cos{\theta}$ where the angle $\theta$ is measured relative to the normal. Scattering surfaces that satisfy Lambert's cosine law scatter phonons isotropically regardless of their incident angle. Generally diffusive scattering has been found to be a good model for phonon-surface reflections, likely due to some small roughness in the surface.

Phonons are strictly-speaking eigenstates of a Hamiltonian that describes an infinitely large, periodic lattice. This description necessarily breaks down at the detector surfaces and there is some probability that phonons down convert to lower energy daughters at the surface. The details of this process would be highly material / surface treatment dependent but the probability of this occurring for a particular phonon-surface interaction will be small for a high purity crystal. It is generally easiest to tune this probability by running numerous Monte Carlo to find the best value. In the CDMS detectors we have measured a loss of $\sim$0.1\% for each phonon-surface interaction~\cite{Leman2011}.

It is the goal of the experimental setup to absorb the phonons into some sensor and provide instrumentation into a data acquisition system. The details for the phonon-sensor interaction probability are again complicated and highly depend on the type of absorber and attachment / fabrication details. Acoustic mismatch theory provides a good starting point for analytic calculations and can be performed over both normal and non-normal incidence angles. The relevance of these calculations can be lost however when detector to detector variations are considered. Additionally the angular dependence of such calculations can be washed out when integrating over a distribution of phonon incident angles and phonon energies. In practice it is usually again easiest to tune this probability by running numerous Monte Carlo and identifying the best fit value. In the CDMS detector there is additionally an amorphous silicon dielectric in between the crystal and thin metal films; we have empirically matched a phonon-aSi-aluminum interaction (details of the interaction are discussed in Section~\ref{sec:QPDC}) probability of $\sim$33\% with the remaining 67\% diffusively scattering back into the crystal~\cite{Leman2011, Leman2011_4}.  For any well designed detector, phonon absorption into the instrumentation sensors will dominate over other loss processes allowing the probability to be tuned by matching pulse decay times.

\subsection{Time Steps}
It would be grossly inefficient to run all phonons with the same
time step. Therefore we generate scattering times according to the
distributions in Sections~\ref{sect:IsotopeScatter}
and~\ref{sect:AnharmonicDecay}. The scattering and decay
probabilities go like $P = 1- \exp(-t / \tau)$ where $\tau$ is a combination of isotope scattering and anharmonic decay rates as given by Matthiessen's rule $1/ \tau = 1/\tau_{iso} + 1/\tau_{anh}$.

This scattering time has to be compared with the time that the phonon will take to interact with a surface and the event that occurs soonest will be chosen for each individual phonon. If it is determined that the bulk interaction time is less than the surface interaction time,  then it must be determined which event occurs based on their relative, frequency dependent rates. This can be done by drawing a uniform random number $u$ and comparing the rate of the process in question to the total rate. For example, if $u < (1/ \tau_{anh}) / (1/ \tau_{anh} + 1/ \tau_{iso})$ then an anharmonic decay is selected.

\subsection{Random Number Sampling}

Only in rare circumstances will a uniform random number $u$ be needed in Monte Carlo without some transformation. Often we are trying to draw the number $x$ out of the probability distribution function (PDF) $f$. An efficient method for transforming $u$ to the desired probability distribution function $f$ is to integrate $f$ to find the cumulative distribution function $F = \int f$. The cumulative distribution function (CDF) $F$ has the desirable property that it is bounded by $[0,1]$ as is $u$. The CDF $F$ is then inverted and solved at $u$ to determine $f$,  $x = F^{-1}(u)$~\cite{NumericalRecipes}. 

As an example, we first considering the bulk interaction rate where the probability of having an interaction is $P = 1 - \exp(-t / \tau)$.  After integration and inversion, the randomly generated scattering time is given by
\begin{equation}
t_{\text{scatter}} = -\tau \ln(u).
\end{equation}

As a second example we can consider diffuse scattering off the detector walls. In a spherical coordinate system where $\theta$ represents the angle from the surface normal, the PDF in this coordinate is $f = \sin(\theta)$. This is also easily integrated and inverted to yield $F^{-1} = \arccos(u)$. Care must be taken in this example however since $\arccos$ is defined over the domain [-1,1] which modifies to $F^{-1} = \arccos(2u-1)$.

There are times when a PDF cannot be analytically integrated to yield a CDF or the CDF cannot be inverted. This is an unfortunate situation since an expensive rejection technique is required. This technique involves drawing a pair of uniform random numbers $[u_x, u_y]$ where $\min(x) \leq u_x \leq \max(x)$ and $\min(f) \leq u_y \leq \max(f)$. If $u_y < f(u_x)$ then $u_x$ is retained, otherwise $u_x$ is rejected and the process is repeated. The inefficiency of this method is related to the area coverage $c = \int f(x) dx / (\max(f)\times(\max(x)-\min(x)))$ and $u_x$ will be successfully drawn in one of $c$ attempts.

The rejection method can be improved however for \emph{static} distributions that do not change during the Monte Carlo run. An example includes diffusive scattering off of the side walls of the detector when using a spherical coordinate system. In this case, the $\sin(\theta)$ Jacobian results in the PDF $f = \sin(\theta)^2$ for which $F$ cannot be found analytically. The PDF can be integrated numerically to generate a CDF which is subsequently inverted. The process lacks a certain degree of elegance but is significantly more efficient than using a rejection method.

\subsection{Numerical Constants for Phonon Simulations}

Table~\ref{tab:phononSimConst} lists numerous constants that are used in the phonon simulations. They define the propagation dynamics, scattering and decay rates and energy carrier statistics.

\begin{table}
\caption{Numerical Constants for Phonon Simulations}
\label{tab:phononSimConst}
\begin{tabular}{|p{9.2em}|c|c|c|c|c|}
  \hline
  Parameter Description & Symbol & Units & Silicon & Germanium & Reference \\
  \hline
  \hline
  Isotope Scatter Rate & ($B$) & [s$^3$] & $2.43\times10^{-42}$ & $3.67\times10^{-41}$ & \cite{Tamura1985} \\
  \hline
  Anharmonic Decay Rate & ($A$) & [s$^4$] & $7.41\times10^{-56}$ & $6.43\times10^{-55}$ & \cite{Tamura1985} \\
  \hline
  Longitudinal & ($v_l$) & [m/s] & 9000 & 5310 & \cite{Tamura1985} \\
   Velocity    &         &       &      &      &                   \\
  \hline
  Transverse Velocity & ($v_t$) & [m/s] & 5400 & 3250 & \cite{Tamura1985} \\
  \hline
  Decay Rate Constant & ($\beta$) & & -0.429 & -0.732 & \cite{Tamura1985} \\
  \hline
  Decay Rate Constant & ($\gamma$) & & -0.945 & -0.708 & \cite{Tamura1985} \\
  \hline
  Decay Rate Constant & ($\lambda$) & & 0.524 & 0.376 & \cite{Tamura1985} \\
  \hline
  Decay Rate Constant & ($\nu$) & & 0.680 & 0.561 & \cite{Tamura1985} \\
  \hline
  Density & ($\rho$) & [g/cm$^3$] & 2.33 & 5.32 & \cite{Tamura1985} \\
  \hline
  Decaying Ratio $\frac{L\rightarrow L+T}{All\;Longitudinal\;Decays}$ & & & 0.204 & 0.260 & \cite{Tamura1985} \\
  \hline
  Lattice Constant & ($C_{11}$) & & 1.66 & 1.29 & \cite{Ashcroft} \\
  \hline
  Lattice Constant & ($C_{12}$) & & 0.64 & 0.48 & \cite{Ashcroft} \\
  \hline
  Lattice Constant & ($C_{44}$) & & 0.80 & 0.67 & \cite{Ashcroft} \\
  \hline
  Debye Frequency & ($\nu_{D}$) & [1/s] & $1.5\times10^{13}$ & $0.864\times10^{13}$ & \cite{Ashcroft} \\
  \hline
  Electron-Hole &   & [eV] & 1.17 & 0.75  & \cite{Ashcroft} \\
         Energy &   &      &      &       &                 \\
  \hline
  Mean Electron-Hole Creation Energy &   & [eV] & 3.81 & 2.96  & \cite{Pehl1968} \\
  \hline
\end{tabular}
\end{table}

\section{Quasiparticle Down Conversion}
\label{sec:QPDC}

Phonons have some probability of entering and interacting in the thin aluminum films that are patterned on the surface. These aluminum films make up both the  phonon collecting films and ionization ground lines; it is interactions in the former that are measured in the phonon sensors. If the phonons have energy greater than or equal than twice the superconducting gap $E_{\phi} > 2 E_{\text{gap}} = \Delta$ then Cooper pairs can be broken creating two quasiparticles. These quasiparticles will be of high kinetic energy $E_{\text{kinetic}} = E_\phi - \Delta$ and contain some probability of scattering off phonons. These daughter phonons thereby introducing a population of down converted phonons back into the metal film. These phonons could break additional Cooper pairs in a cascade process that ceases when all phonons have energy below $\Delta$ or have a probability of being reintroduced back into the crystal. The key points in the cascade process are summarized in the following list~\cite{Kaplan1976, Kurakado1982, BrinkThesis}.

\begin{enumerate}

\item Quasiparticle recombination lifetimes are long compared to quasiparticle decay and quasiparticle absorption into the aluminum films and therefore  the recombination processes can be ignored.

\item Quasiparticle decay via absorption of a phonon is suppressed at low temperatures due to a phonon density of states term $n(\Omega)$, where $\Omega$ is the phonon energy, in the Green's function and therefore can be ignored.

\item Quasiparticle decay via emission of a phonon results in phonons with an energy distribution given by $P_{\phi}(\Omega) = \Omega^2 \rho(E-\Omega) \left(1- \frac{\Delta^2}{E (E-\Omega)} \right)$, where $E$ is the quasiparticle energy and the quasiparticle density of states at temperature $T$=0 goes like $\rho(E) = \frac{E}{\sqrt{E^2 - \Delta^2}}$ and $0 \le \Omega \le E$. 

\item Phonons break Cooper pairs producing quasiparticles with an energy distribution given by $P_{\text{QP}}(E) = \left(1+ \frac{\Delta^2}{E(\Omega-E)} \right) \rho(E) \rho(\Omega - E)$, where $\Delta \le E \le \Omega-E$. The phonon is completely absorbed so that the second quasiparticle has energy $E_2 = \Omega-E$.

\item Phonons are lost to the crystal if they reach the aluminum / crystal interface. 

\end{enumerate}

\subsection{Monte Carlo process ordering}
In the physical processes can be ordered as follows

\begin{enumerate}

\item If the phonon energy is sufficient $\Omega > 2 \times E_{\text{gap}}$, a quasiparticle pair is created with the distribution $P_{\text{QP}}(E)$ previously described.

\item If there is a quasiparticle with energy $E \ge 3 \times E_{\text{gap}}$ then the quasiparticle emits a phonon with energy distribution $P_{\phi}(\Omega)$. Quasiparticles with energy $E < 3 \times E_{\text{gap}}$ would shed a phonon with $E \le 2 \times E_{\text{gap}}$ and therefore provide an endpoint for quasiparticle generation. They may be removed from the Monte Carlo.

\item The probability of a phonon escaping the crystal is a function of the distance to reach the aluminum / crystal interface and the phonon / Cooper pair interaction length. Given the large number of phonons it is generally not necessary to track this process in detail and instead a simple model is sufficient. On average, phonons are assumed to populate the center of the aluminum film ($z = l_{\text{Al}}/2$), where $l_{\text{Al}}$ is the aluminum thickness, and the phonons have 1/2 probability of traveling upwards and 1/2 probability of traveling downwards. For the downward going phonons there is an $exp(- (2 \times l_{\text{Al}}/2) / L_0)$ probability of reaching the aluminum / crystal interface before scattering, where $L_0 \sim 720$~nm is a characteristic phonon interaction length~\cite{BrinkThesis}. The factor of $2 \times$ is provided to integrate over different phonon incidence angles. The factor $l_{\text{Al}}/2$ is replaced by $l_{\text{Al}} \times 3/2$ for upward going phonons.

\item If the phonon has not been removed from Monte Carlo in step 2 or reintroduced into the crystal in step 3 then the process repeats at step 1.

\end{enumerate}

\section{Charge Monte Carlo}
\label{sec:chargeHV}
\subsection{Introduction}

Accurate modeling of charge propagation is included in Monte Carlo for numerous reasons. First, the ionization signal, compared to the phonon signal, provides a discriminator between electron-recoil and nuclear-recoil events in the silicon and germanium detectors. Second, electron transport is described by a mass tensor, leading to electron transport which contains components oblique to the applied field. This description is necessary to explain and interpret signals in the primary and guard-ring ionization channels, which function as a fiducial volume cut~\cite{Sasaki1958, Jacoboni1983}. Third, for electron recoils in the germanium bulk, charges drifting through the detector produce a population of Luke phonons which contribute  56\% of the total phonon signal at $\pm$3~V bias. This fraction is understood by considering that for every $E_\text{eh,create}$ of gamma energy, an electron-hole pair is created which contributes $e V_\text{bias}$ of phonon energy; $e V_\text{bias} / (E_\text{eh,create} - E_\text{gap} + e V_\text{bias}) = 56\%$ is the contribution of Luke phonons to the total phonon signal. These phonons' spatial, time, energy and emitted-direction distributions should therefore be properly modeled in Monte Carlo of the detector response. Fourth, phonons created during electron-hole recombination at the surfaces contribute 13\% of the total phonon signal but in a low frequency, ballistic regime that is used to provide a surface-event discriminator. This fraction is understood by considering that for every $E_\text{eh,create}$ of gamma energy, $E_\text{gap}$ of phonon energy is released at the surface; $e V_\text{gap} / (E_\text{eh,create} + e V_\text{bias}) = 13\%$ is the contribution of electron-hole gap energy phonons to the total phonon signal. These phonons also need to be properly modeled in a Monte Carlo. 
 
\begin{figure}
\begin{center}
\includegraphics[width=7cm, bb=200 50 500 500]{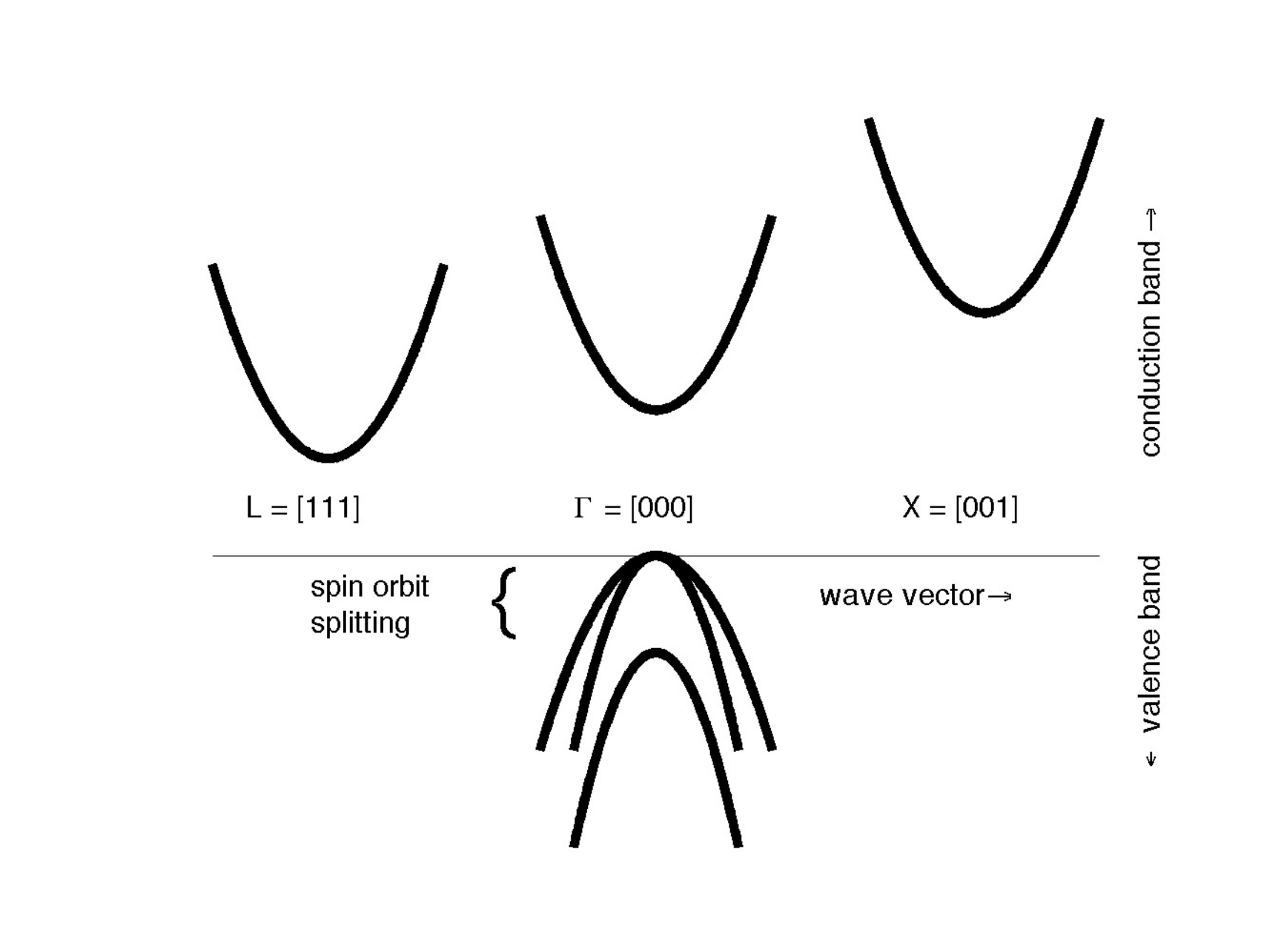}
\end{center}
\caption[Band structure] { \label{fig:bandGe} Energy band structure of germanium, showing the L-valleys at $\langle 111 \rangle$, the $\Gamma$ valley at $\langle 000 \rangle$, and the X-valley at $\langle 100 \rangle$. Symmetry results in 8 total L-valleys and 6 total X-valleys.}
\end{figure}

Germanium has an anisotropic band structure described schematically in Figure~\ref{fig:bandGe} and shows energy band structure in which the hole ground state is situated in the $\Gamma$ band's [000] direction and the electron ground state is in the L-band [111] direction. Hole propagation dynamics are relatively simple due to propagation in the $\Gamma$ band and the isotropic energy-momentum dispersion relationship $\epsilon(\overrightarrow{k}) = \hbar^2 k^2 / 2 m$. Electron propagation dynamics are significantly more complicated due to the band structure and anisotropic energy-momentum relationship. At low fields and low temperatures, electrons are unable to reach sufficient energy to propagate in the $\Gamma$ or X-bands, and are not considered necessary to consider in Monte Carlo. The electron energy-momentum dispersion relationship is given by $\epsilon(\overrightarrow{k}) = (\hbar^2 / 2) \times  (k_{\parallel}^2 / m_{\parallel} + k_{\perp}^2 / m_{\perp} )$, where the longitudinal and transverse mass ratio $m_{\parallel} / m_{\perp} \sim 19.5$.

This chapter will proceed by describing hole propagation and scattering, electron propagation and scattering utilizing a Herring-Vogt transformation and finally electron-hole recombination. Higher order mass terms and scattering processes which occur at high electric fields are discussed elsewhere in the literature~\cite{AubryFortuna2010}.

\subsection{Holes: propagation and scattering with isotropic bands and isotropic phonon velocity}

Hole propagation dynamics are described by momentum evolution in an electric field $\hbar \frac{d\overrightarrow{k}}{dt} = e \overrightarrow{E}$ and propagation in position space $\overrightarrow{v} = \hbar \overrightarrow{k} / m_c$, where $m_c$ is the effective carrier mass.

\begin{figure}
\begin{center}
\includegraphics[width=12cm, bb=0 0 1033 361]{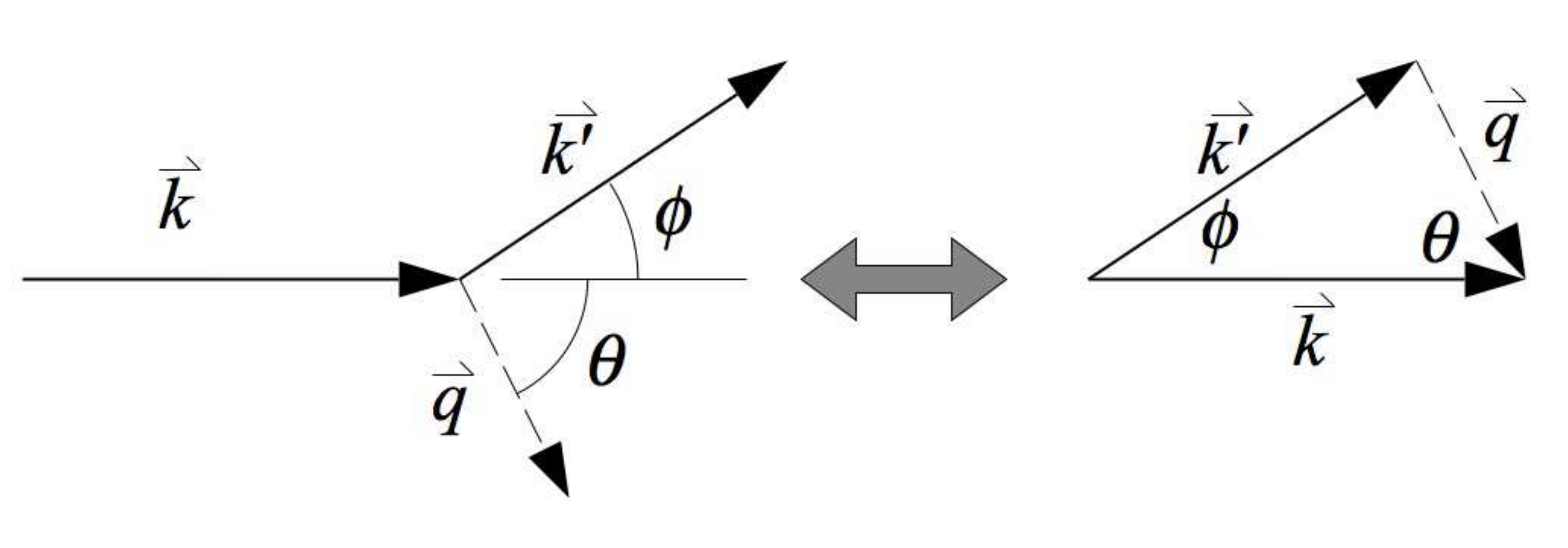}
\end{center}
\caption{ \label{fig:scatter} Charge carrier with initial wave vector $\vec k$ and final wavevector $\vec k'$ scattering off of lattice at angle $\phi$ with respect to $\vec k$ and emitting a phonon with wavevector $\vec q$ at angle $\theta$ with respect to $\vec k$, where the vector momenta sum as shown on right.}
\end{figure}

Charge carriers cannot accelerate indefinitely however, and the shedding of Neganov-Luke phonons~\cite{Neganov1985, Luke1988} limits their speed to around the longitudinal phonon phase velocity $v_{L}$. As described in Figure~\ref{fig:scatter}, charge-phonon scattering is an elastic processes, conserving both momentum $\overrightarrow{k} - \overrightarrow{k'} = \overrightarrow{q}$ (where $\overrightarrow{k}$ and $\overrightarrow{k'}$ are the initial and final hole momentum vectors and $\overrightarrow{q}$ is the phonon momentum vector) and energy $\epsilon - \epsilon' = \hbar \omega$ (where $\epsilon$ and $\epsilon'$ are the initial and final hole energies and $\hbar \omega$ is the phonon energy). The phonon energy-momentum dispersion relationship is given by $\omega = v_{L} q$. Due to the low carrier energy, Umklapp processes~\cite{Kittel} in which $\overrightarrow{k} - \overrightarrow{k'} = \overrightarrow{q} + \overrightarrow{G}$, where $G$ is a reciprocal lattice vector, are suppressed.

Energy-momentum conservation coupled with the previous dispersion relationships leads to the final states $k'^2 = k^2 + q^2 - 2kq \cos(\theta)$ and $q = 2(k \cos{\theta} - k_L)$ and 

\begin{equation}
\cos(\phi) = \frac{k^2 - 2k_s(k \cos{\theta}-k_s) - 2(k \cos(\theta) - k_s)^2}{k \sqrt{(k^2 - 4 k_s (k \cos(\theta)-k_s)}},\\
\end{equation}

\noindent where $\theta$ is the angular displacement between $\overrightarrow{k}$ and $\overrightarrow{q}$, $\phi$ is the angular displacement between$\overrightarrow{k}$ and $\overrightarrow{k'}$, and $k_L$ is defined as $k_L = m v_L / \hbar$. If we can determine a scattering rate $\tau$ and phonon angular displacement $\theta$, then we can use these formulae to find the final states.

Fermi's Golden Rule provides the transition probability per unit time per unit energy as 

\begin{equation}
P_{k,k' \pm q} = \frac{2 \pi V}{ \hbar}  \left|< \overrightarrow{k} \pm \overrightarrow{q} |H| \overrightarrow{k}> \right|^2 \delta(E-E' \mp \hbar \omega).
\end{equation}

For phonon emission processes, 
\begin{equation}
\left|< \overrightarrow{k} \pm \overrightarrow{q} |H| \overrightarrow{k'}> \right|^2 = \frac{C^2 \hbar}{2 V \rho v_L}q (n_q +1),
\end{equation}

\noindent where $C$ is the deformation potential constant and $n_q$ is the phonon occupation number given by $n_q = 1/ \left(e^{\hbar \omega / k_B T} - 1 \right)$. A characteristic length can be defined as $l_0 = \frac{\pi \hbar^4 \rho}{2 m^3 C^2}$. The transition probability can be integrated over $\theta$ and $E'$ to obtain a scattering rate 

\begin{equation}
1/ \tau = \frac{1}{3} \frac{v_L}{l_0} \frac{k}{k_L} \left[ 1 - \frac {k_L}{k}   \right]^3.
\end{equation}

The angular distribution then follows to be 

\begin{equation}
P(k,\theta) d\theta= \frac{v_L}{l_0} \left(\frac{k}{k_L}\right)^2 \left(\cos(\theta) - \frac{k_L}{k} \right)^2   \sin{\theta} d\theta
\end{equation}

 \noindent where $0 \leq \theta \leq \arccos( k_l/k ) < \pi/2$.

The phonon scatter azimuthal rotation angle is uniformly distributed about $0 \leq \vartheta \leq 2 \pi$. The charge carrier azimuthal rotation angle is required to be $\varphi = \pi + \vartheta$. 

It is critical that the time steps in the Monte Carlo are sufficiently small that the scattering rates are approximately constant during a time step. A method for ensuring this is discussed in Sections~\ref{ChargeTimeStep1} and ~\ref{ChargeTimeStep2}.

\subsection{Electrons: propagation and scattering with anisotropic bands and isotropic phonon velocity}

Electron propagation and scattering is complicated by the anisotropy of the electron bands but can be simplified by performing first a transformation into a space defined by the vectors $(\overrightarrow{e_{\parallel}},\overrightarrow{e_{\perp 1}},\overrightarrow{e_{\perp 2}})$ (where $\overrightarrow{e_{\parallel}}$ is aligned with [111] and the other two are perpendicular) and then a Herring-Vogt transformation into a space where the electron bands are isotropic~\cite{Herring1956, Jacoboni1983}. The Herring-Vogt transformation is non-unitary and in the $(\overrightarrow{e_{\parallel}},\overrightarrow{e_{\perp 1}},\overrightarrow{e_{\perp 2}})$ space is given by

\begin{equation} 
\mathcal{T} = T_{HV} =  \left(
  \begin{array}{ccc}
    \sqrt{\frac{m_c}{m_\parallel}} & 0 & 0 \\
    0 & \sqrt{\frac{m_c}{m_\perp}} & 0 \\
    0 & 0 & \sqrt{\frac{m_c}{m_\perp}} \\
  \end{array}
\right),
\end{equation}

\noindent but the speed of sound $v_L$ remains unchanged and isotropic.

In this space, $k^* = \mathcal{T} k$, $\epsilon = \frac{\hbar^2 {k^*}^2}{2 m}$, $\hbar \frac{d\overrightarrow{k^*}}{dt} = e \overrightarrow{E^*}$, and the effective mass is given by $3/m_c = 1/m_{\parallel} + 2/m_{\perp}$. The change in velocity, in position space, is found by a back transform, incorporating both the mass and momentum transforms, and is given by $ v=\frac{\hbar}{m} \mathcal{T} k^*$.

After the Herring-Vogt transform is used to find the electric-field induced acceleration, the electrons shed phonons via the same prescription given to the holes. The phonon and electron momentum is found first in the Herring-Vogt space and the back transform $\mathcal{T}^{-1}$ is applied to return to position space. The only additional concern is correctly back transforming the phonon momentum due to the non-unitarity of the Herring-Vogt transform. To handle this, we maintain the phonon momentum magnitude (ie, conserve energy), but use the back transform $\mathcal{T}^{-1}$ to find the correct angular distribution.

\subsection{Charge Time Steps, First Order}
\label{ChargeTimeStep1}

Determining an \emph{efficient} time step size, $dt$ for charge transport is complicated since the scattering time $\tau$ varies as the charge carrier accelerates. In this charge transport model it was decided to use sufficiently small $dt$ such that $\tau$ is relatively constant at each iteration. The requirement, \emph{sufficiently small}, in practice means that energy is conserved in the transport process and the following is a detailed description of how this requirement is implemented.

By observing the charge Monte Carlo over numerous field strengths and in germanium, it was observed that scattering limits the maximum possible charge momentum magnitude to about

\begin{equation}
\label{eq:elStep}
k_{max,el} =  13 \times  k_L    \left| \overrightarrow{E} \right| ^{1/3} 
\end{equation}
\begin{equation}
\label{eq:hStep}
k_{max,h} =  6.8 \times k_L   \left| \overrightarrow{E} \right| ^{1/3}
\end{equation}

\noindent for electrons and holes respectively. These momenta $k$ are then used to determine stepping times which conserve energy; the shed Luke phonon energy must equal the change in potential energy $e \Delta V$ at the carrier drifts. Again running numerous Monte Carlo it is determined that a stepping time of $dt = \tau / 2$ is sufficiently small to conserve energy. Larger stepping times will result in a deficit of Neganov-Luke phonons being created.

\subsection{Charge Time Steps, Second Order}
\label{ChargeTimeStep2}
The earlier described first order method can be improved upon by developing a second order method. The challenge is to efficiently and accurately determine the time until a charge sheds a Neganov-Luke phonon, which is challenging due to the changing interaction time, as the charge is accelerated by the field. An inverse CDF technique would be advantageous and one is developed here that adapts to the changing interaction time.

This is done by sampling the scattering rate at two different times $\tau_0$ and $\tau_1$. We start with the differential equation $\frac{dN}{dt} = -a_0 N$ and expand to next order in time
\begin{equation}
\frac{dN}{dt} = (-a_0 -a_1 t) N .
\end{equation}
Integrating, we can obtain
\begin{equation}
\ln N = -(a_0 t + a_1  t^2 /2 ) .
\end{equation}

\noindent
This continues with the standard technique of solving for the CDF and inverting to obtain $a_1 t^2 /2 + a_0 t + \ln(u) = 0$ which can be solved for the scattering time $t = \frac{ -a_0 \pm \sqrt{a_0^2 - 2 a_1 \ln{u}}}{a_1}$. The positive root is retained as physical which provides a scattering time of 

\begin{equation}
t = \frac{\sqrt{a_0^2 - 2 a_1 \ln{u}} - a_0}{a_1} .
\end{equation}

\noindent
This is completed by recognizing that 

\begin{equation}
a_0 = \left. \frac{dN}{dt} \right|_{t=t_0} = \tau_0^{-1}
\end{equation}

and 

\begin{equation}
a_1 = \frac{1}{t_1 - t_0} \left( \left. \frac{dN}{dt} \right|_{t=t_1}  -  \left. \frac{dN}{dt} \right|_{t=t_0}  \right)     =    \frac{1}{t_1 - t_0}   ( \tau_1^{-1} - \tau_0^{-1} ).
\end{equation}

There is a maximum sampling time step ($t_1-t_0$) that can be used before the linear interpolation of scattering rates is inaccurate. As in the first order case, this results in lack of energy conservation. It is found that the electron sampling time step can be a factor of $~\sim$15 greater than the time step shown in Equation~\ref{eq:elStep} and the hole sampling time step by a factor of $\sim$20 greater than the time step in Equation~\ref{eq:hStep}. Given these sampling time steps, most Neganov-Luke phonons are produced in a time $t < t_1$, hence this method is much more efficient than the first order method. The efficiency of this second order method also implies that pursuit of higher order methods will not yield much additional improvement in computational efficiency.

This method also couples well to a second order spatial transport method. The velocity form of the Verlet algorithm~\cite{Verlet1967, Swope1982} is convenient and given by a description that should look familiar.

\begin{enumerate}

\item $x(t + \Delta t) = x(t) + v(t) \Delta t + \frac{1}{2} a(t) \Delta t^2$

\item $v(t + \Delta t) = v(t) + \frac{1}{2} (a(t) +a(t + \Delta t)) \Delta t$

\end{enumerate}

This can be easily modified to incorporate the second order inverse CDF sampling method via the following procedure:

\begin{enumerate}

\item Make a guess for the step size $\Delta t$, which we will call $\Delta t_0$

\item $x(t + \Delta t_0) = x(t) + v(t) \Delta t_0 + \frac{1}{2} a(t) \Delta t_0^2$

\item $\tau_0 = \tau_0(v(t))$

\item $v(t + \Delta t_0) = v(t) + \frac{1}{2} (a(t) +a(t + \Delta t_0)) \Delta t_0$

\item $\tau_1 = \tau_1(v(t + \Delta t_0))$

\item From second order inverse CDF method, determine the randomly distributed scattering time $\Delta t_1$

\item If $\Delta t_1 < \Delta t_0$

	\begin{enumerate}
	
	\item $x(t + \Delta t_1) = x(t) + v(t) \Delta t_1 + \frac{1}{2} a(t) \Delta t_1^2$
	
	\item $\tau_2 = \tau_1(v(t + \Delta t_1))$
	
	\item $v(t + \Delta t_1) = v(t) + \frac{1}{2} (a(t) +a(t + \Delta t_1)) \Delta t_1$
	
	\item Save $\tau_2$ and $a(t+\Delta t_1)$ for use in next iteration
	
	\end{enumerate}

\item Else, save $\tau_1$ and $a(t+\Delta t_0)$ for use in next iteration

\end{enumerate}

Since holes are described by a scalar mass, this procedure is straightforward. There is however a slight modification to this procedure that is useful for electrons. Due to the use of the Herring Vogt transform, it is generally easier to keep track of momentum in the Herring Vogt space, $k^*$, rather than velocity. We also identify that the acceleration is given by $a = \mathcal{X}^{-1} \frac{\mathcal{T} \mathcal{T} }{m_c} \mathcal{X} E$, where the transform $\mathcal{X}$ from the cartesian space to the space defined by basis vectors $(\overrightarrow{e_{\parallel}},\overrightarrow{e_{\perp 1}},\overrightarrow{e_{\perp 2}})$ is shown explicitly.

\subsection{Select constants for charge Monte Carlo}

\begin{table}[ht]
  \begin{center}
 \caption{Physical constants for Si and Ge crystals. The isotropic hole effective mass $m_h$, and the anisotropic electron effective masses $m_\parallel$ and $m_\perp$ are $\parallel$ and $\perp$, respectively, to the conduction valley axes, and conductivity effective mass $3/m_c = 1/m_\parallel + 2/m_\perp$. The incident energy per final electron-hole pair is $\epsilon_{eh}$, $v_L$ the speed of sound, and $l_0 = {\pi \hbar^4 \rho}/{(2 m^3 \Xi^2)}$ is the characteristic range for carrier scattering where $\Xi_1$ (from~\cite{Jacoboni1983}) or $\Xi_{\text{ fit}}$ (fit to data~\cite{Sundqvist2009}) is the deformation potential.}
  \label{tab:constants}
  \vspace{.1in}
  \begin{tabular}{|c|c|c|c|c|}
    \hline
    &\multicolumn{2}{|c|}{Silicon}& \multicolumn{2}{|c|}{Germanium}\\\hline
    & Electrons & Holes & Electrons & Holes \\\hline
    $m_h/m_e$           &  -    & 0.5   &   -   & 0.35  \\\hline
    $m_{\parallel}/m_e$ & 0.91 &  -   & 1.58  &  -  \\\hline
    $m_{\perp}/m_e$     & 0.19  &  -  & 0.081 &  -  \\\hline
    $m_c/m_e$           & 0.26  &   -   & 0.12  &   -   \\\hline
    $v_L$ (km/s)      & \multicolumn{2}{|c|}{9.0} & \multicolumn{2}{|c|}{5.4} \\\hline
    $\rho$ (g/cm$^3$)  & \multicolumn{2}{|c|}{2.335} & \multicolumn{2}{|c|}{5.323} \\\hline
    $\Xi_1$ (eV)        & 9.0  & 5.0  & 11.0  & 4.6  \\\hline
    $\Xi_{\text {fit}}$ (eV)        &  -  &  -   & 11.0  & 3.4  \\\hline
    $l_0$ ($\mu$m)        & 16.9  &  7.5  & 257  & 108 \\\hline
  \end{tabular}
  \end{center}
 \end{table}

\section{Electric-Field Lookup}
\label{BCC}

\subsection{Electric-Field Lookup from Triangulated Mesh}

A numerical electric field model is necessary for the charge transport described in Section~\ref{sec:chargeHV}.  The simplest model is a constant, longitudinally directed field. However it may be desirable to include fringing fields and details from the electrode structure. A more accurate model will utilize a triangulated mesh. A 3-d mesh contains nodes, with each mesh node containing an associated electric potential $V$. At points other than a mesh node, the potential $V$ must be interpolated. The MATLAB programming language offers a few options for this interpolation, the fastest of which, utilizing a barycentric-coordinates linear interpolation via the \emph{TriScatteredInterp} class. The Computational Geometry Algorithms Library (CGAL) is available for C{}\verb!++!~\cite{cgal} though this paper will be presented for a MATLAB implementation. Furthermore, the barycentric transformation involves a linear transformation which implies that the electric field is constant within a triangulation, a property which can be exploited to speed up computation.

MATLAB's method of looking up the potential, while very convenient, is not efficient considering the number of repeated field queries that occur before the carrier has moved to a location with a differing field. Efficiency can be improved by exploiting the fact that a charge remains within its triangle for numerous iterations and that the field is constant with the triangulation. On the contrary, MATLAB solves for the potential at every iteration in its lookup procedure, which is a bit slow. These repeated searches can be avoided but at the expense of significant code complexity. The effort is justified however as charge transport imposes a dominant computational expense in Monte Carlo. 

\subsection{Barycentric Coordinates}

Given a triangulation (the mesh is made of tetrahedra in 3-space but the term triangulation often persists) with four node points $\mathbf{r}_1, \mathbf{r}_2, \mathbf{r}_3$, and $\mathbf{r}_4$, the arbitrary point $\mathbf{r}$ can be described by the barycentric coordinates $\lambda_1, \lambda_2, \lambda_3$, and $ \lambda_4$ where 

\begin{equation}\label{eq:bccdef}
\mathbf{r} = \lambda_1 \mathbf{r}_1 + \lambda_2 \mathbf{r}_2 + \lambda_3 \mathbf{r}_3 + \lambda_4 \mathbf{r}_4.
\end{equation}

\noindent The additional constraint is imposed that 

\begin{equation}\label{eq:lambdaconstr}
\lambda_1 + \lambda_2 + \lambda_3 + \lambda_4 = 1.
\end{equation}

\begin{figure}
\begin{center}
\includegraphics[width=10cm, bb=0 0 300 200]{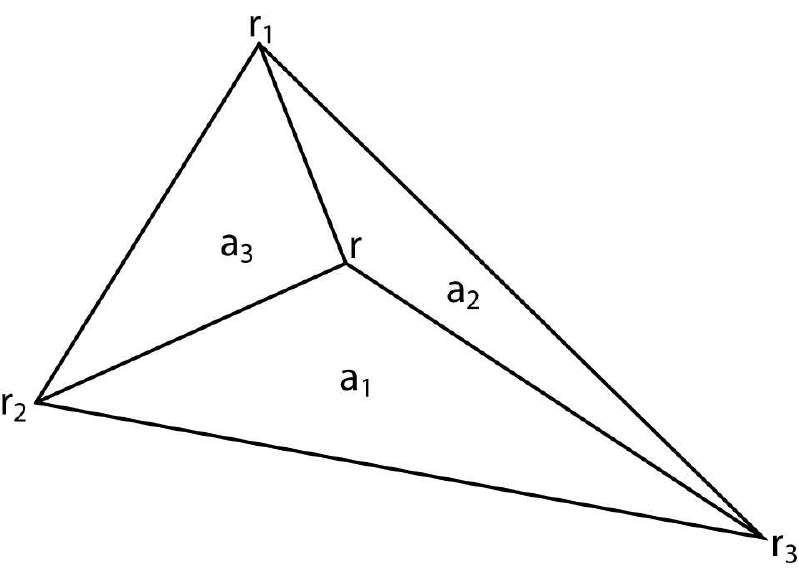}
\end{center}
\caption[Area Components] { \label{fig:AreaComponents}
Mesh node points $\mathbf{r}_1, \mathbf{r}_2$, and $\mathbf{r}_3$ along with the probe point $\mathbf{r}$. The areas $a_1, a_2$ and $a_3$ are identically equal to the barycentric coordinates $\lambda_1, \lambda_2$ and $\lambda_3$.}
\end{figure}

The barycentric coordinates $\lambda$ become more intuitive when thought of as area (volume in 3-d) coordinates (see Figure~\ref{fig:AreaComponents}). In this paradigm, consider the 2-d node points $\mathbf{r}_1, \mathbf{r}_2$, and $\mathbf{r}_3$ along with the probe point $\mathbf{r}$. To start with, let's normalize the area enclosed by $\mathbf{r}_1, \mathbf{r}_2$, and $\mathbf{r}_3$ to $1$ (this normalization is equivalent to the constraint $\sum_i \lambda_i = 1$). Then we can consider the three different areas enclosed by 1) $\mathbf{r}_2, \mathbf{r}_3$, and $\mathbf{r}$ ($a_1$), 2) $\mathbf{r}_1, \mathbf{r}_3$, and $\mathbf{r}$ ($a_2$) and 3) $\mathbf{r}_1, \mathbf{r}_2$, and $\mathbf{r}$ ($a_3$). It turns out that these areas ($a_1$, $a_2$ and $a_3$) are identically equal to the barycentric coordinates $\lambda_1, \lambda_2$, and $\lambda_3$, providing a quick and intuitive interpretation of the barycentric coordinates. The process and interpretation is the same in 3-d when volume is substituted for area. It is not actually recommended to calculate $\lambda$ through this procedure but to instead follow the procedure in Section~\ref{sec:bccMath}.

\subsection{Barycentric Coordinate Formulae}
\label{sec:bccMath}

In this section we derive formulae useful for solving the barycentric coordinates and electric-potential. We start again with the definitions given by equations~\ref{eq:bccdef} and~\ref{eq:lambdaconstr}. After separating equation~\ref{eq:bccdef} into the $\mathbf{x}, \mathbf{y}$, and $\mathbf{z}$ components we can solve for the $\lambda$ through the following linear procedures and the formula is written out explicitly below.

\begin{equation}
T =  \left(
  \begin{array}{ccc}
    x_1 - x_4   &   x_2 - x_4 & x_3 - x_4\\
    y_1 - y_4   &   y_2 - y_4 & y_3 - y_4\\
    z_1 - z_4   &   z_2 - z_4 & z_3 - z_4\\
  \end{array}
\right).
\end{equation}

\begin{equation}\label{eq:bccPracticalDef}
\left(
  \begin{array}{c}
    \lambda_1     \\
    \lambda_2    \\
    \lambda_3   \\
  \end{array}
\right) = T^{-1} (r-r_4) = T^{-1} \left(
  \begin{array}{c}
    x-x_4   \\
    y-y_4   \\
    z-z_4   \\
  \end{array}
\right),
\end{equation}

\noindent where $\lambda_4 = 1 - (\lambda_1 + \lambda_2 + \lambda_3 )$. Explicitly we can write out

\begin{equation}
T^{-1} =  \frac{1}{ \text{det}(T)} \left(
  \begin{array}{c}
    \hspace{-30 mm} (y_2 - y_4)(z_3-z_4) - (y_3 - y_4)(z_2 - z_4)   \\
        \hspace{0 mm} (z_2 - z_4)(x_3-x_4) - (z_3 - z_4)(x_2 - x_4)   \\
      \hspace{30 mm} (x_2 - x_4)(y_3-y_4) - (x_3 - x_4)(y_2 - y_4)\\
       \\
    \hspace{-30 mm} (y_3 - y_4)(z_1-z_4) - (y_1 - y_4)(z_3 - z_4)    \\
        \hspace{0 mm} (z_3 - z_4)(x_1-x_4) - (z_1 - z_4)(x_3 - x_4)    \\
      \hspace{30 mm} (x_3 - x_4)(y_1-y_4) - (x_1 - x_4)(y_3 - y_4)\\
       \\
    \hspace{-30 mm} (y_1 - y_4)(z_2-z_4) - (y_2 - y_4)(z_1 - z_4)    \\
        \hspace{0 mm} (z_1 - z_4)(x_2-x_4) - (z_2 - z_4)(x_1 - x_4)    \\
      \hspace{30 mm} (x_1 - x_4)(y_2-y_4) - (x_2 - x_4)(y_1 - y_4)\\
  \end{array}
\right).
\end{equation}

\noindent where $
 \text{det}(T) = ...\\(x_1-x_4) \times [(y_2-y_4)(z_3-z_4) - (y_3-y_4)(z_2-z_4)]   +  ... \\
(x_2-x_4) \times [(y_3-y_4)(z_1-z_4) - (y_1-y_4)(z_3-z_4)] + ...\\
 (x_3-x_4) \times [(y_1-y_4)(z_2-z_4)-(y_2-y_2)(z_1-z_4)]
 $.
 
 The potential is simple to solve for in the barycentric coordinate system and equal to $V = \lambda_1 V_1 + \lambda_2 V_2 + \lambda_3 V_3 + \lambda_4 V_4$. This procedure is more obvious when one considers the connection to area coordinates.

 \subsection{Barycentric Coordinate Procedures and Shortcuts}
 
The procedure for finding which tetrahedra a charge resides is not unique and a canned MATLAB procedure (or CGAL library in C{}\verb!++!) can be utilized for this step. After the tetrahedra in which the carrier resides is determined, the electric field, $\mathbf{E} = - \mathbf{\nabla} V$ is computed. This can be performed by probing the potential at four locations ($\mathbf{r}, \mathbf{r}+\delta\mathbf{x}, \mathbf{r}+\delta\mathbf{y}$ and $\mathbf{r}+\delta\mathbf{z}$) and computing gradients. The drawback is that this procedure requires conversion to barycentric coordinates and electric potential lookup for three additional points. These steps can be eliminated with some simple derivations that are outlined below. Most of these steps provide a conceptual framework and only the last step is actually computed.

First we consider two points $\mathbf{r'}$ and $\mathbf{r''} = \mathbf{r'} + \delta \mathbf{x}$ and find their associate barycentric coordinates $\lambda'$ and $\lambda''$. 

 \begin{equation}
\left(
  \begin{array}{c}
    \lambda'_1     \\
    \lambda'_2    \\
    \lambda'_3   \\
  \end{array}
\right)  = T^{-1} \left(
  \begin{array}{c}
    x'-x_4   \\
    y'-y_4   \\
    z'-z_4   \\
  \end{array}
\right).
\end{equation}
 
\noindent and

 \begin{equation}
\left(
  \begin{array}{c}
    \lambda''_1     \\
    \lambda''_2    \\
    \lambda''_3   \\
  \end{array}
\right)  = T^{-1} \left(
  \begin{array}{c}
    x''-x_4   \\
    y''-y_4   \\
    z''-z_4   \\
  \end{array}
\right),
\end{equation}

\noindent where, as always, $\lambda_4$ is constrained by $\sum \lambda = 1$. Next we solve for the potentials $V'$ and $V''$

\begin{equation}
V' = \lambda'_1 V_1 + \lambda'_2 V_2 + \lambda'_3 V_3 + \lambda'_4 V_4 = \lambda' \mathbb{V}
\end{equation}

\noindent and

\begin{equation}
V'' = \lambda''_1 V_1 + \lambda''_2 V_2 + \lambda''_3 V_3 + \lambda''_4 V_4 = \lambda'' \mathbb{V},
\end{equation}

\noindent where $\mathbb{V} = (V_1, V_2, V_3, V_4)$.

It follows that $E_x = (V' - V'') / \delta x$ but another route is preferable. We can define $\delta \lambda_{1,2,3} = {\lambda'}_{1,2,3} - {\lambda''}_{1,2,3} = - T^{-1} \left(
  \begin{array}{c}
    \delta x   \\
    0   \\
    0   \\
  \end{array}
\right)$.

The fourth element in $\delta \lambda$ is determined by remembering that $\sum \lambda' = \sum \lambda'' = 1$ it follows that $\sum \delta \lambda = 0$ and therefore $\delta \lambda_4 = - (\delta \lambda_1 + \delta \lambda_2 + \delta \lambda_3)$. We can define a matrix related to $T^{-1}$ but of size $4 \times 3$ to make future calculations easier,

\begin{equation}\label{eq:modT}
\mathbb{T}^{-1} = \left(
  \begin{array}{c c c}
 {T^{-1}}_{1,1} & {T^{-1}}_{1,2} & {T^{-1}}_{1,3} \\
 {T^{-1}}_{2,1} & {T^{-1}}_{2,2} & {T^{-1}}_{2,3} \\
 {T^{-1}}_{3,1} & {T^{-1}}_{3,2} & {T^{-1}}_{3,3} \\
   -\sum_i{{T^{-1}}_{i,1}}   & -\sum_i{{T^{-1}}_{i,2}} & -\sum_i{{T^{-1}}_{i,3}}   \\
  \end{array}
\right).
\end{equation}

We have now obtained a convenient and efficient method for calculating $E_x$
\begin{equation}
E_x = \frac{1}{\delta x} \mathbb{V} \delta \lambda =   \frac{1}{\delta x} \mathbb{V} \mathbb{T}^{-1} \left(
  \begin{array}{c}
    \delta x   \\
    0   \\
    0   \\
  \end{array}
\right) = 
 \mathbb{V} \mathbb{T}^{-1} \left(
  \begin{array}{c}
    1   \\
    0   \\
    0   \\
  \end{array}
\right)
\end{equation}

\begin{equation}\label{eq:eField}
= V_1 {\mathbb{T}^{-1}}_{1,1} + V_2 {\mathbb{T}^{-1}}_{2,1} + V_3 {\mathbb{T}^{-1}}_{3,1} + V_4 {\mathbb{T}^{-1}}_{4,1}
\end{equation}
and similarly for $E_y$ and $E_z$. Equations~\ref{eq:modT} and~\ref{eq:eField} are used in Monte Carlo and the fields are reused until the charge leaves the triangle boundaries.

\subsection{Determining if a Charge Leaves a Triangle}

The assumption that is exploited in these computations is that the charge remains in a particular triangle for many iterations. This is true, especially in the detector bulk where mesh nodes are sparse, but eventually the charge will enter a different triangle. With the concept of area components, it is clear that one of the barycentric coordinates $\lambda$ will be zero when the charge is at a surface. By continuation, the same $\lambda$ will become negative when the charge leaves the triangle. This can be tested efficiently via the linear calculations already derived in Equation~\ref{eq:bccPracticalDef}. When a charge enters a new triangle then $T, \mathbb{T}$ and $\mathbf{E}$ must be recalculated and stored to memory.

\section{Transition Edge Sensor Simulations}
\label{ch:TESSim}

\subsection{Introduction}
Energy collected in the aluminum fins diffuses through the aluminum fins and is collected into the tungsten Transition Edge Sensors (TESs)~\cite{KIrwin1995, Irwin2005} with an efficiency of $\sim$10\%. The aluminum quasiparticle fins combined with the tungsten TESs together are referred to as Quasiparticle-trap-assisted Electrothermal-feedback Transition-edge-sensors (QET)~\cite{Cabrera2000}. The quasiparticle downconversion process was described previously in Section~\ref{sec:QPDC} and here we focus on modeling of the TES sensor, which provides the electrical signal which is readout in the experiment. Compared to the phonon simulations, the TES simulations are a bit simpler; most of the electrical and thermal processes that make up the simulation are likely to be more obvious to the reader. 

\begin{figure}
\begin{center}
\includegraphics[height=18cm, bb=0 0 734 1482]{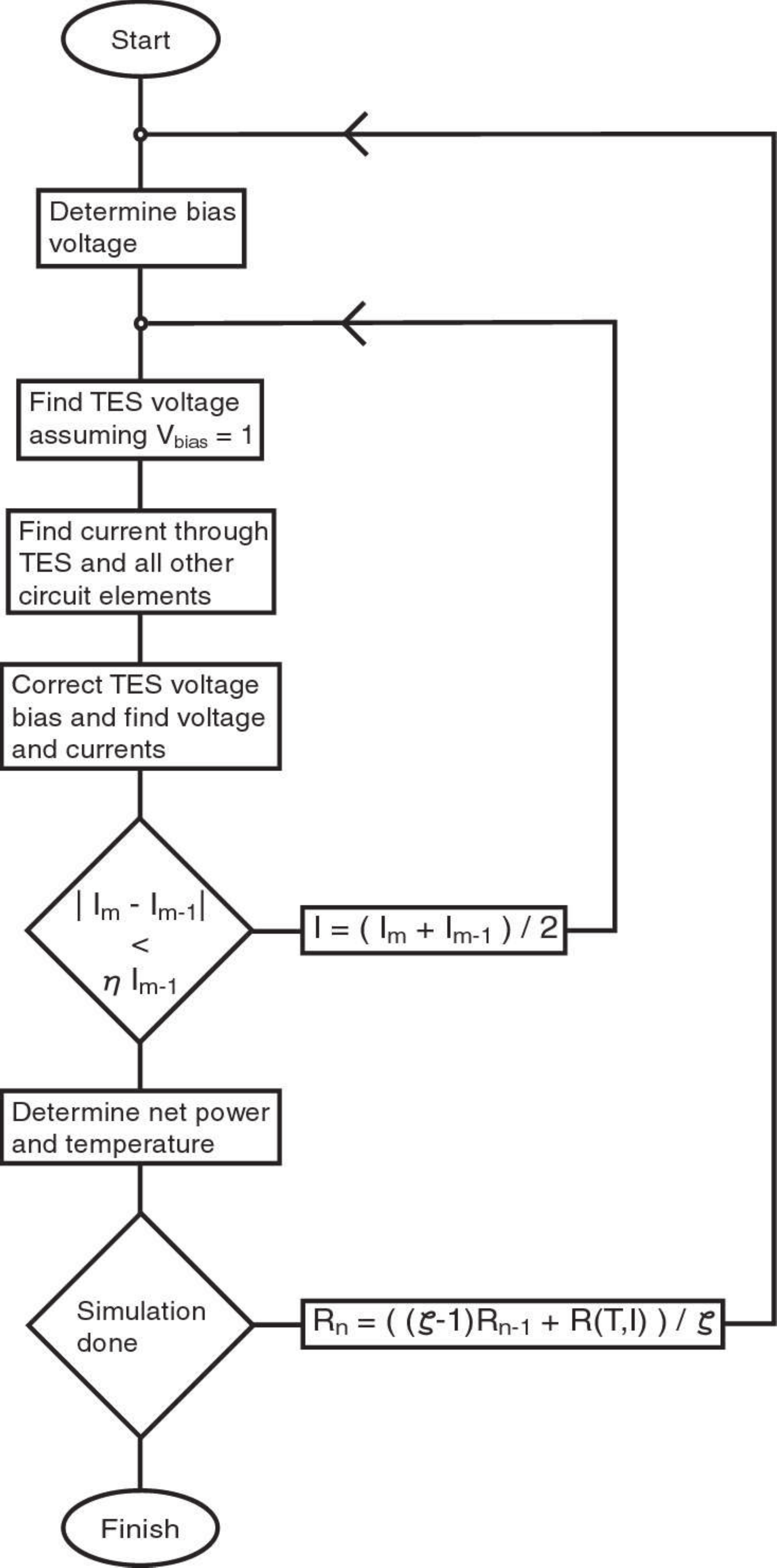}
\end{center}
\caption[TES Simulation Flowchart] { \label{fig:TESSimFlowchart} TES
simulation flowchart.}
\end{figure}

There are, however, tricky numerical issues in solving for the TES voltages. A description of the process follows, and a flowchart (see Figure~\ref{fig:TESSimFlowchart}) helps in understanding it. First, the biasing conditions have to be determined. When a TES is run as a phonon sensor (they are also found in many x-ray detectors), the bias voltage is held constant. However, current--voltage (IV) characterization curves are often studied in which the bias current is swept through a range of values. The biasing circuitry and TES resistance $R = R(T,I)$, where $T$ is temperature and $I$ is current across some physical extent, are modeled. The TES electron temperature is affected by three processes: 1) Joule heating warms the TES; 2) Conduction into the substrate cools it; 3) Diffusion through the TES spreads heat. It is of course the goal to run the simulation both quickly and accurately.

\subsection{Electrical Circuit Modeling}

\begin{figure}
\begin{center}
\includegraphics[width=10cm, bb=0 0 535 219]{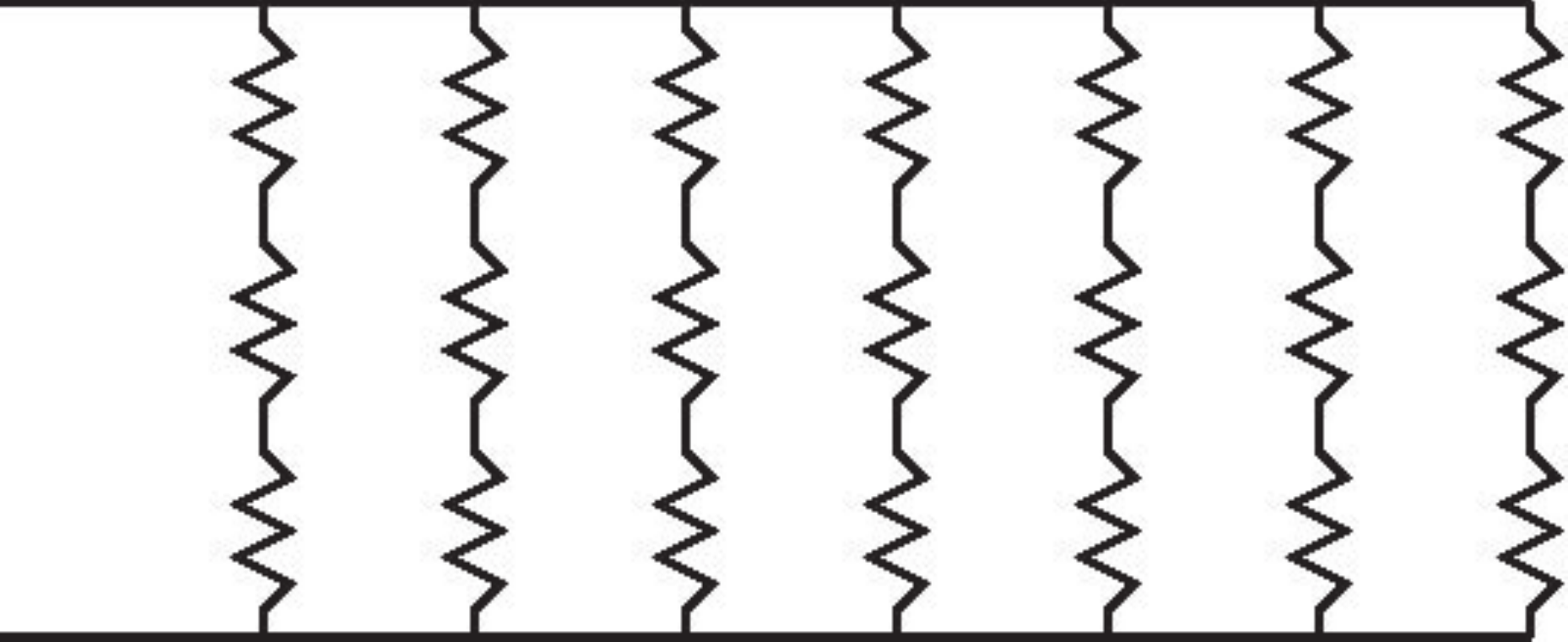}
\end{center}
\caption[TES Resistor Interconnects / Finite Elements] {
\label{fig:resistorInterconnects} TES resistor interconnects as
modeled using a finite element approximation.}
\end{figure}

The TESs are relatively simple to model as each can be thought of as a one dimensional object (see Figure~\ref{fig:resistorInterconnects}). Generally, a TES is modeled with a minimum of two nodes and as many as $\sim$100 if it is desired to reproduce normal-superconducting phase separation~\cite{LemanThesis, Anderson2011, Anderson2011_2}. The functional form of the resistance at each node has generally been given by

\begin{equation}
R_{i,j} = \frac{R_{max} - R_{min}}{2}   \left\{ 1+  \tanh\left[\frac{T_{i,j} -
T_c \left(1- \frac{|I_{i,j}|}{I_c}\right)^{n_{sc}}  }{T_w} \right]   \right\}, 
\end{equation}
\noindent for $I_{i,j}$ small and $R_{i,j} = R_{max}$, for $I_{i,j}$ large;
where $R_{max}$ is the TES's resistance in the normal state, $R_{min}$ is the superconducting resistance (numerically it is best to not set $R_{min}$ to zero but rather some small value $\sim
10^{-8}~\Omega$), $T_c$ is the midpoint temperature of the superconducting transition, $T_w$ is the temperature width of the superconducting transition, $I_c$ is the critical current (see below), and $n_{sc} \sim 2/3$ is motivated by~\cite{Tinkham, Irwin1998}.

The critical current is provided by Ginzburg-Landau theory and near $T_c$ is equal to $I_c = 3.52 \sqrt{\frac{k_B C_n}{\hbar R_n}} T_c \left(1- \frac{T}{T_c}\right)^{3/2}$, where $k_B$ is the Boltzmann constant, and $C_n$ is the heat capacity. Special care needs to be taken when calculating the critical current when modeling TESs that are wired in parallel. Implicit in the Ginzburg-Landau equation is that \emph{one} thin film superconductor contains the current while in CDMS detectors, the TES is distributed over $n_{\text{TES}}$ TESs. This cause the heat capacity $C_n$ to scale down by $1/n_{\text{TES}}$ and $I_c$ scales down by $1/ \sqrt{n_{\text{TES}}}$ for each of the TESs. There is an offsetting factor when the currents from all the $n_{\text{TES}}$ TES which causes $I_c$ to scale up by $n_{\text{TES}}$. The end result is that $I_c$ for the entire collection of $n_{\text{TES}}$ parallel TESs scales like $\sqrt{n_{\text{TES}}}$.

\begin{figure}
\begin{center}
\includegraphics[width=15cm, bb=0 0 1200 900]{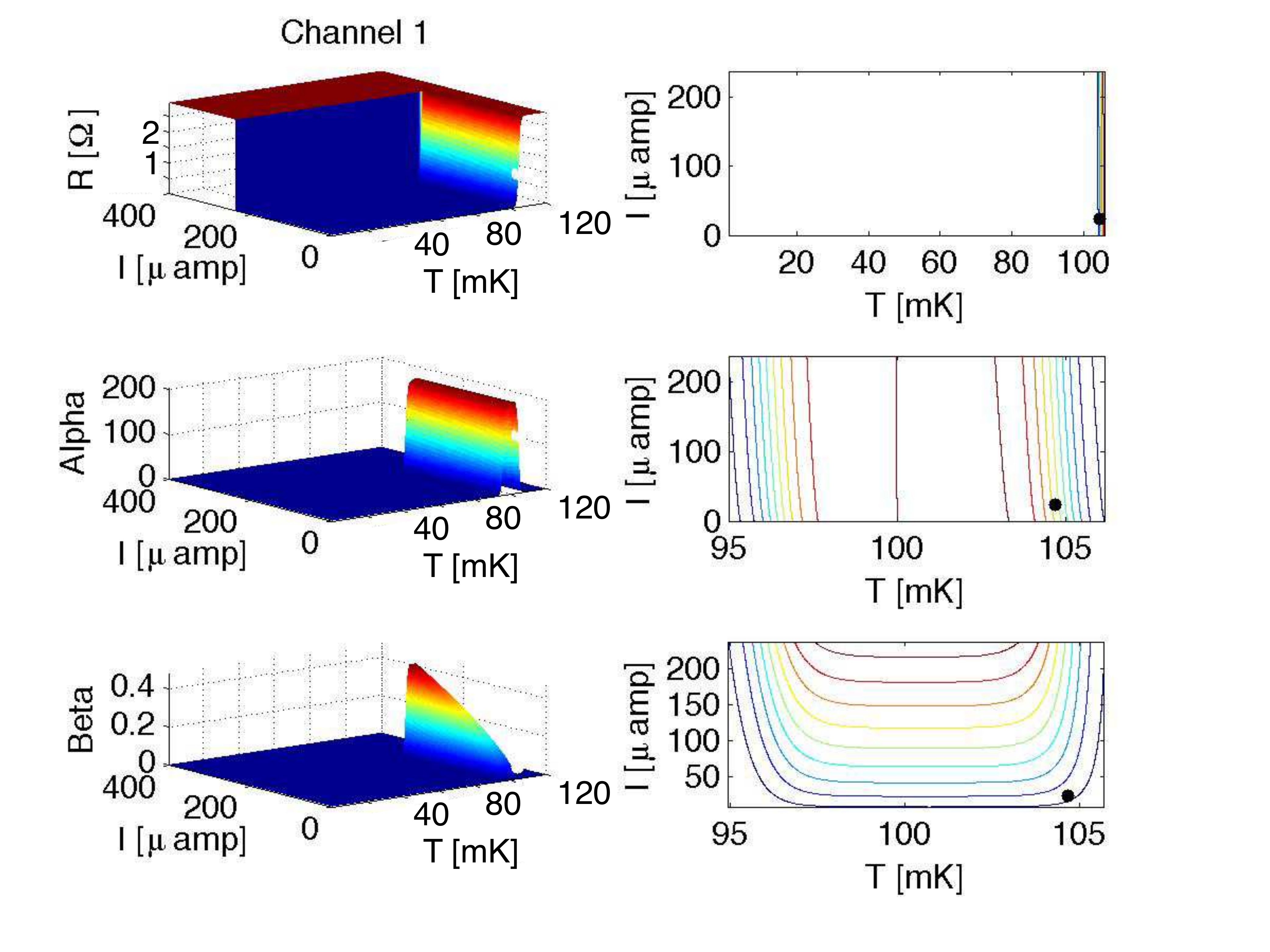}
\end{center}
\caption[R = R(T,I) for channel 1] { \label{fig:RTI_chan1} Surface and contour plots of $R = R(T,I)$, $\alpha = \frac{dR}{dT}\frac{T}{R} = \alpha(T,I)$ and $\beta = \frac{dR}{dI}\frac{I}{R} = \beta(R,I)$ for a high-Tc, inner iZIP channel. The colors in the surface plot indicate the value of resistance, alpha and beta with blue representing 0 and red the highest value in the figure. The contour plots show the same information but over a limited current and temperature region. The black dot indicates a nominal bias region, which will affect noise and pulse shape after a radiation interaction in the detector. The gradient in resistance and temperature is generally along the temperature direction, whereas for $\beta$ it is in a mixed $-T +I$ direction.}
\end{figure}

The requirement that the resistance $R=R_{max}$ at large current $I$ is required for IV curves that sweep from the superconducting to normal states. There is no precise definition of large $I$, but it is chosen to be much higher than the quiescent operating current while low enough that it is not too much higher than the normal to superconducting transition current. A plot of $R=R(T,I)$, $\alpha = dR/dT \times T/R$ and $\beta = dR/dI \times I/R$ is shown in Figure~\ref{fig:RTI_chan1} .

\begin{figure}
\begin{center}
\includegraphics[width=10cm, bb=0 0 767 849]{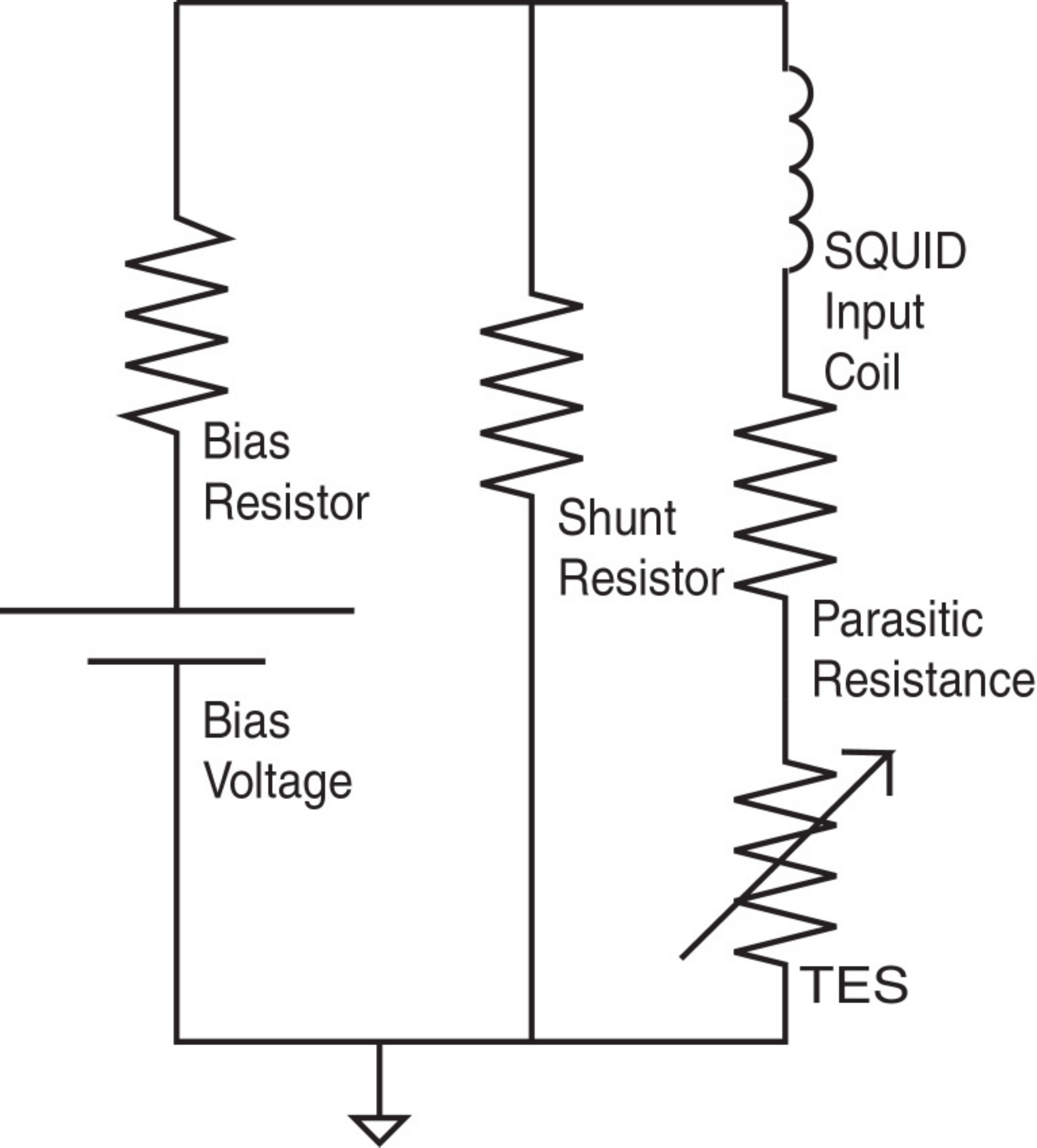}
\end{center}
\caption[TES Simulation Biasing Circuitry] {
\label{fig:TESBiasCircuit} TES simulation biasing circuitry.
Modeling reflects the biasing circuitry.}
\end{figure}

Modeling of the TES and biasing circuitry (see Figure~\ref{fig:TESBiasCircuit}) is complicated by our desire to be able to create IV curves in which the bias voltage ($V_{bias}$) is swept from zero to some large value and back to zero again. This difficulty comes about from the current dependence on resistance and also due to the sharp change in current that occurs during the transition from the superconducting to normal states. This difficulty can be overcome if we use a damping factor to prevent rapid resistance changes. To determine the TES resistance at step $n$ we let
\begin{equation}
R_n = \frac{(\zeta-1)R_{n-1} + R(T,I)}{\zeta}
\end{equation}
where $\zeta$ is a stabilization constant whose value $\sim 5$ is found by trial and error. Picking a large value makes current transitions numerically stable, but requires slower step sizes $dt$ to retain accuracy.

Next, we solve for the voltages at each TES node (to be discussed below) and determine the resulting voltages and currents. We also model the inductor as discussed below. It is likely that the currents that we solve for differ from those that we originally used to compute $R = R(T,I)$. I keep evaluating $R_m(T,I_{m})$, $V$ and $I$ letting
\begin{equation}
I_{m}= \frac{I_{m-1} + I_{m-2}}{2}
\end{equation}
until
\begin{equation}
|I_m - I_{m-1}| < \eta I_{m-1}
\end{equation}
where $\eta \sim 10^{-3}$ is satisfied for all nodes, ensuring a self consistent answer. I have enumerated the individual steps $m$ here to prevent confusion with the steps $n$ that involve simulation time steps $dt$ (again, see Figure~\ref{fig:TESSimFlowchart}).

The rest of the circuit, including the inductor, is modeled by finding the macroscopic current through the TES $I_n$. A simple circuit analysis reveals that it is given by {\large{
\begin{equation}
I_{n} = \frac{\frac{V_{bias}R_{shunt}}{R_{bias}+R_{shunt}} + \frac{L
I_{n-1}}{dt}}{ \frac{R_{bias}R_{shunt}}{R_{bias}+R_{shunt}} +
R_{true} + R_{parasitic} + \frac{L}{dt}}.
\end{equation}}}

\subsubsection{TES Voltage Modeling}
In the simulation, temperature, and therefore resistance, is modeled at each node. However for the purpose of modeling the electrical circuit the interconnects need to have a resistance which is given by averaging the resistance of the two nodes that they connect.

We will now set up the system of equations $\mathbf{G} V = B$ where the conductance matrix, $\mathbf{G}$, is a sparse, square matrix and $B$ is a vector that describes the electrical boundary conditions . The form of the conductance matrix $\mathbf{G}$ is simplified significantly if we treat the individual TESs as one dimensional object. In general (regardless of the 1-d simplification) nodes either are voltage biased (which includes grounding) or are not. Those that are not biased (or grounded) impose current conservation requirements. From Kirchhoff's current law, the equation to be satisfied for nodes that have four adjacent nodes is

\begin{equation}
\frac{V_{\text{left}} - V}{R_{\text{left}}} + \frac{V_{\text{right}} - V}{R_{\text{right}}} + \frac{V_{\text{top}} - V}{R_{\text{top}}} + \frac{V_{\text{bottom}} - V}{R_{\text{bottom}}}= 0.
\end{equation}
Nodes without four adjacent nodes will result in the appropriate
terms being removed, for example a node with a top, left and right
neighbors is given by
\begin{equation}
\frac{V_{\text{left}} - V}{R_{\text{left}}} + \frac{V_{\text{right}} - V}{R_{\text{right}}}  + \frac{V_{\text{top}} - V}{R_{\text{top}}}= 0.
\end{equation}
The result on the $\mathbf{G}$ matrix is straightforward. The
diagonal elements are given by
\begin{equation}
\mathbf{G}_{ii} = -\frac{1}{R_{\text{left}}} -\frac{1}{R_{\text{right}}}.
\end{equation}
The non-diagonal elements that couple adjacent nodes are given by
\begin{equation}
\mathbf{G}_{\text{terms which couple adjacent nodes}} =
\frac{1}{R_{\text{adjacent}}}.
\end{equation}

Nodes that have their voltage fixed either through voltage biasing or grounding satisfy the equation $V = V_{\text{bias}}$. They are given the coupling terms
\begin{equation}
\mathbf{G}_{ii} = 1,
\end{equation}
the other terms in the row are
\begin{equation}
\mathbf{G}_{ij} = 0,
\end{equation}
where $i \neq j$.

Now we can set up $B$ and complete Kirchhoff's current law. For the rows in $\mathbf{G}$ that describe current conservation,
\begin{equation}
B_i = 0.
\end{equation}
Whereas the rows in $\mathbf{G}$ that describe voltage biasing,
\begin{equation}
B_i = V_{\text{bias}}.
\end{equation}
We can then solve
\begin{equation}
V = \mathbf{G}^{-1} B,
\end{equation}
to find the voltages at all TES nodes.

\subsection{Thermal Processes}

There are many thermal processes which affect the TES temperature: the addition of heat from phonons, Joule heating, cooling to the substrate and thermal diffusion within the TES. 
\begin{equation}
P = P_\text{QP} + P_\text{Joule} + P_\text{substrate} + P_\text{diffusion}
\end{equation}
Most of these processes are relatively simple to model and discussed only briefly. Diffusion however is more involved and needs to be considered if it is necessary to properly describe normal-superconducting phase separation in the TES. 

\subsubsection{Phonon Heat}
The process of phonon -- quasiparticle down conversion was described previously in Section~\ref{sec:QPDC} as a process in which phonons energy is removed from the crystal. This process results in a population of quasiparticles (broken Cooper pairs) which diffuses through the aluminum films. In the CDMS detectors, these aluminum films are coupled to the tungsten TESs. Due to the large number of quasiparticles (the aluminum quasiparticle gap is $\sim$100~$\mu$eV, small compared to the 10-100~keV gamma energy) the details of where individual phonons interact in the film average out and we can use average quantities. In the CDMS detectors, the measured efficiency of quasiparticles reaching the TES is $\varepsilon_{\text{QP}}\sim$10\%. The remaining $\sim$90\% are trapped in the aluminum (details are discussed in references~\cite{Saab2000, Pyle2004}.
\begin{equation}
P_{\text{QP}} = \varepsilon_{\text{QP}} P_{\text{phonon}}
\end{equation}

\subsubsection{Joule Heating}
After determining the voltages at each node, Joule heating is
calculated by
\begin{equation}
P_J = \frac{V^2}{R}.
\end{equation}

\subsubsection{Substrate Cooling}
Power into the substrate is given by
\begin{equation}
P_S = -\Sigma \text{ Vol } \left(T_{\text{TES}}^n - T_{\text{substrate}}^n \right)
\end{equation}
where $\Sigma$ is the electron-phonon coupling parameter, Vol is the TES volume and $n$ is an exponent that describes electron-phonon coupling.

\subsubsection{Diffusion Within the TES}
Thermal diffusion is simplified by assuming that the QETs are thermally decoupled from each other due to poor conduction through the substrate. The process is then described by a 1-d diffusion process
\begin{equation}
\frac{\partial T_\text{diffusion}}{\partial t} = D \triangle T,
\end{equation}
where
\begin{equation}
\triangle T = \nabla \cdot \nabla T = \frac{\partial^2T}{\partial
x^2} 
\end{equation}
and $D$ is given by $K/C_V$ where $K$ is the thermal conductivity and $C_V$ is the heat capacity per unit volume~\cite{Kittel}.

For non-boundary nodes, $\triangle T$ can be approximated by considering the midpoint and two adjoining points. To see this let us first consider just one dimension, specifically the term $\frac{\partial^2T}{\partial x^2}$. First we compute $\frac{\partial T}{\partial x}$. In the discrete approximation we can consider some point $T_i$ and either its right neighbor or left neighbor $T_{i+1}$ or $T_{i-1}$, where $i$ represents the position of some node in the $x$ direction. We obtain either
\begin{equation}
\left.\frac{\partial T}{\partial x}\right|_{\text{right}} \rightarrow
\frac{T_{i+1} - T_i}{\delta x}
\end{equation}
or
\begin{equation}
\left.\frac{\partial T}{\partial x}\right|_{\text{left}} \rightarrow
\frac{T_{i} - T_{i-1}}{\delta x},
\end{equation}
where $\rightarrow$ denotes the move from continuous to discrete space. Next we compute
\begin{align}\label{eq:intNodeDiff}
\frac{\partial}{\partial x} \frac{\partial T}{\partial x}
\rightarrow \frac{\left.\frac{\partial T}{\partial x
}\right|_{\text{right}} - \left.\frac{\partial T}{\partial
x}\right|_{\text{left}}}{\delta x}
\notag\\
\notag\\
= \frac{T_{i-1} - 2T_i + T_{i+1}}{(\delta x)^2}.
\end{align}

Boundary nodes are computed differently. To conserve energy we impose the Neumann boundary condition
\begin{equation}
\left.\nabla T\right|_{\text{boundary}} = 0.
\end{equation}
Making the discrete approximation is simple, we compute
\begin{equation}\label{eq:extNodeDiff}
T_{i} - T_{i-1} = 0
\end{equation}
for boundary nodes.

$T$ will be the temperature matrix of size $n \times m$ where $n$ and $m$ represents the nodes in the $y$ and $x$ directions respectively. It would be nice to convert the following procedures for computing $\triangle T$ into a matrix operation which considering Equations~\ref{eq:intNodeDiff} and~\ref{eq:extNodeDiff} is fortunately is quite simple. The matrix $\tilde{\triangle}_x$ is of size $m \times m$ and allows discrete calculations of $\triangle T$. Specifically,
\begin{equation}
\frac{\partial^2T}{\partial x^2} \rightarrow \frac{T
\tilde{\triangle}_x}{(\delta x)^2}
\end{equation}
where $\tilde{\triangle}_x$ has the
form
\begin{equation} \label{eq:del2}
 \left(
  \begin{array}{cccccccc}
    -1 & 1 & 0 & 0 & \cdots & 0 & 0 & 0\\
    1 & -2 & 1 & 0 & \cdots & 0 & 0 & 0\\
    0 & 1 & -2 & 1 & \cdots & 0 & 0 & 0\\
      & & & & \ddots & & &   \\
    0 & 0 & 0 & 0 & \cdots & 1 & -2 & 1 \\
    0 & 0 & 0 & 0 & \cdots & 0 & 1 & -1 \\
  \end{array}
\right).
\end{equation}
This form is equivalent to Equations~\ref{eq:intNodeDiff} and~\ref{eq:extNodeDiff} and that the signs are correct for the boundary nodes to ensure heat moves from hot to cold nodes.

The diffusion algorithm is then given by
\begin{equation}
T_{n+1} = T_n + D \;\delta t \left( \frac{T \tilde{\triangle}_x}{(\delta
x)^2} \right).
\end{equation}
The algorithm should be stable as long as the Courant--Friedrichs--Lewy condition
\begin{equation}
\delta t < \frac{(\delta x)^2}{2 D}
\end{equation}
is satisfied~\cite{Courant1928, Burden} but the Joule heating and substrate cooling terms result in a less stable algorithm and the stability is observed to become poorer as the TES cell size is reduced. The onset of instability is due to local temperature deviations due to small numerical imprecision. These deviations can be recognized by comparing the maximum absolute temperature difference in adjacent cells $T_{\text{diff,adj}} = \text{max}( | (T_{i+1} - T_{i} | )$ to the maximal temperature difference in the TES $T_{\text{diff,max}} = \text{max}(T) - \text{min}(T)$. The number of time steps is then scaled up by  $T_{\text{diff,adj}} / T_{\text{diff,max}} \times N / f$, where $N$ is the number of cells along the TES and $f$ is a factor $\sim 10$ to reduce the instability caused by the inclusion of Joule heating and substrate cooling.

\subsection{Parsing Phonons Into The TESs}
One issue that needs to be considered before running a TES simulator is the parsing of phonons into the multiple QETs which are wired in parallel. It is generally more computationally efficient to ignore the exact location with and without aluminum coverage and to use an average approximation. In this case, we need only determine the physical region of each QET structure. This can be approximated by assuming each QET to have the same area. For a TES channel with radial and angular extent $R_1$, $R_2$, $\Theta_1$ and $\Theta_2$ and containing $N$ discrete circuit elements, the area of each element is simply 

\begin{equation} \label{eq:QETarea1}
a = \frac{1}{2 N} \left(R_2^2-R_1^2 \right) \left(\Theta_2-\Theta_1 \right).
\end{equation}

\noindent For the first ring of TES elements we could also consider the area of each element to be 

\begin{equation} \label{eq:QETarea2}
a = \frac{1}{2} \delta \theta [(R_1 + \delta r)^2 - R_1^2] = \frac{\delta\theta}{2} (2 R_1 \delta r + \delta r^2),
\end{equation}

\noindent where $\delta r$ and $\delta \theta$ are the elements' size in the radial and angular directions. Equating the two descriptions of area we get

\begin{equation} \label{eq:QETdrdt}
\frac{1}{2 N}  (R_2^2-R_1^2)(\Theta_2-\Theta_1) = \delta \theta \left(R_1 \delta r + \frac{1}{2} \delta r^2 \right).
\end{equation}

We can now impose a proportionality between the element's linear dimensions in the radial and theta direction (measured at $r = R_1 + \delta r/2$) then we get the relationship $\delta r = c (R_1 + \delta r/2) \delta \theta$, where $c$ is a proportionality constant. This can be substituted into Equation~\ref{eq:QETdrdt} to eliminate $\delta \theta$ and we obtain 
\begin{equation}
N ~\delta r^3 + 2 R_1 N ~\delta r^2 - \frac{2 c}{4} (R_2^2-R_1^2)(\Theta_2-\Theta_1) ~\delta r - \frac{2 c}{2} (R_2^2-R_1^2)(\Theta_2-\Theta_1) R_1= 0.
\end{equation}

\noindent This equation can be solved for $\delta r$ and provides one real root. 

If $\delta r > R_2 - R_1$ then the lower value $R_2-R_1$ is of course selected otherwise a $\delta r$ for each ring is proposed to be equal and rounded off to $(R_2 - R_1) / \text{round} ((R_2 - R_1) / \delta r)$. The round is included to ensure an integer number of rings. The number of $\theta$ divisions $n$ for this ring is simply

\begin{equation}
n = \text{round} ((R_1 + \delta r)^2 - R_1^2  (\Theta_2 - \Theta_1) / 2 / a),
\end{equation}

\noindent where the number of elements is further constrained by $n \geq 1$. 

This solution isn't really self consistent since we have imposed too many completing desires, namely that the radial and angular dimensions satisfy a proportionality condition and that there are an integer number of elements in a ring. The former can be relaxed, but the integer number of elements is mandatory. We can obtain a solution with dimensions close to the desired proportionality condition if we set $\delta\theta = (\Theta_2-\Theta_1) / n$ and  $\delta r = \sqrt{2a/\delta\theta + R_1^2} - R_1$. After this first ring is built up the procedure repeats, with $R_1 \rightarrow R_1-\delta r$ until the entire channel area has been segmented.

\subsection{Numerical Constants for TES Simulation}

Table~\ref{tab:TESConstants} lists several constants related to the TES thermodynamic properties and quasiparticle collection efficiency from aluminum to tungsten.

\begin{table}[ht]
  \begin{center}
 \caption{Physical constants for tungsten TES and aluminum fin simulation, from reference~\cite{Anderson2011_2}. }
  \label{tab:TESConstants}
  \vspace{.1in}
  \begin{tabular}{|c|c|c|c|}
    \hline
    
    name & symbol & value & units \\\hline 
    \hline
    $n$                   & electron-phonon coupling exponent & 5 & n/a    \\\hline
    $\Sigma$         &  electron-phonon coupling constant  & 4.8$\times$10$^8$   & W m$^{-3}$ K$^{-5}$  \\\hline
    $C$                   & heat capacity                                          & 37                                 & J  m$^{-3}$  K$^{-1}$    \\\hline
    $D$                   & diffusion constant                                  & 4$\times$10$^{-4}$   & m$^2$ s                               \\\hline
    $\varepsilon_{\text{QP}}$  & QP detection efficiency                        & 10\% &   n/a                     \\\hline

  \end{tabular}
  \end{center}
 \end{table}


\section{Final Remarks}

This paper has covered many physics and Monte Carlo topics in cryogenic radiation-detectors that utilize phonon and ionization readout. The bulk response is different for gamma and neutron interactions, the former producing a larger ratio or ionization energy. Propagation of both the phonons and charge carriers is anisotropic, the phonons transport dependent on dispersion relations and the charge carrier transport dependent on mass tensors. Phonon interactions are complicated at the surface by electron-phonon downconversion requiring additional Monte Carlo effort. The TES readout simulation is relatively straightforward relying on basic thermal and electrical processes. Various numerical techniques including PDF sampling and tricks for improving efficiency have been discussed.

Monte Carlo of detectors is a valuable tool for detector design, characterization and data analysis. Often counter intuitive results can be quickly discovered in Monte Carlo which aids in prioritizing laboratory R\&D efforts. Furthermore, Monte Carlo can help to explain features in data, or at least rule out models. 

Detailed Monte Carlo studies in CDMS~\cite{Leman2011, Leman2011_2, Leman2011_3, Leman2011_4, McCarthy2011, Anderson2011, Anderson2011_2} and EDELWEISS~\cite{Broniatowski2004, Broniatowski2011} detectors can be found in the literature. These studies include discussion of quantities which are relevant in the tuning and their effects on phonon pulse shape, charge transport, TES phase-separation and TES noise.

With the power of Monte Carlo and increased use of low temperature cryogenic detectors, a general package is becoming more appealing. The GEANT4 collaboration is beginning to implement phonon and charge transport modules into the GEANT4 toolkit and will expand the benefits of this research to others outside the cryogenic detector community~\cite{Brandt2011}.

\section*{Acknowledgments}

I would like to thank the entire CDMS collaborators for valuable discussions over the years, especially Adam Anderson, Paul Brink, Blas Cabrera, Enectali Figueroa-Feliciano, Scott Hertel, Peter Kim, Kevin McCarthy, Matt Pyle, Bernard Sadoulet, Kyle Sundqvist and Betty Young. BC generously provided his note on ``Electron-Phonon Scattering'', which was adapted for this review. KM has been an invaluable contributor to the CDMS Monte Carlo studies cited in the references and PK was invaluable in tirelessly running Monte Carlo on the SLAC computing farm.



\begin{thebibliography}{10}

\bibitem{Ahmed2010}
{ {CDMS II Collaboration}; {Ahmed}, Z.; {Akerib}, D.~S.; {Arrenberg}, S.;
  {Bailey}, C.~N.; {Balakishiyeva}, D.; {Baudis}, L.; {Bauer}, D.~A.; {Brink},
  P.~L.; {Bruch}, T.; {Bunker}, R.; {Cabrera}, B.; {Caldwell}, D.~O.; {Cooley},
  J.; {Cushman}, P.; {Daal}, M.; {DeJongh}, F.; {Dragowsky}, M.~R.; {Duong},
  L.; {Fallows}, S.; {Figueroa-Feliciano}, E.; {Filippini}, J.; {Fritts}, M.;
  {Golwala}, S.~R.; {Grant}, D.~R.; {Hall}, J.; {Hennings-Yeomans}, R.;
  {Hertel}, S.~A.; {Holmgren}, D.; {Hsu}, L.; {Huber}, M.~E.; {Kamaev}, O.;
  {Kiveni}, M.; {Kos}, M.; {Leman}, S.~W.; {Mahapatra}, R.; {Mandic}, V.;
  {McCarthy}, K.~A.; {Mirabolfathi}, N.; {Moore}, D.; {Nelson}, H.; {Ogburn},
  R.~W.; {Phipps}, A.; {Pyle}, M.; {Qiu}, X.; {Ramberg}, E.; {Rau}, W.;
  {Reisetter}, A.; {Saab}, T.; {Sadoulet}, B.; {Sander}, J.; {Schnee}, R.~W.;
  {Seitz}, D.~N.; {Serfass}, B.; {Sundqvist}, K.~M.; {Tarka}, M.; {Wikus}, P.;
  {Yellin}, S.; {Yoo}, J.; {Young}, B.~A. and {Zhang}, J. } {\em Science} {\bf
  327}, 2010.

\bibitem{Ahmed2011}
{ {Ahmed}, Z.; {Akerib}, D.~S.; {Arrenberg}, S.; {Bailey}, C.~N.;
  {Balakishiyeva}, D.; {Baudis}, L.; {Bauer}, D.~A.; {Brink}, P.~L.; {Bruch},
  T.; {Bunker}, R.; {Cabrera}, B.; {Caldwell}, D.~O.; {Cooley}, J.; {Do Couto E
  Silva}, E.; {Cushman}, P.; {Daal}, M.; {Dejongh}, F.; {di Stefano}, P.;
  {Dragowsky}, M.~R.; {Duong}, L.; {Fallows}, S.; {Figueroa-Feliciano}, E.;
  {Filippini}, J.; {Fox}, J.; {Fritts}, M.; {Golwala}, S.~R.; {Hall}, J.;
  {Hennings-Yeomans}, R.; {Hertel}, S.~A.; {Holmgren}, D.; {Hsu}, L.; {Huber},
  M.~E.; {Kamaev}, O.; {Kiveni}, M.; {Kos}, M.; {Leman}, S.~W.; {Liu}, S.;
  {Mahapatra}, R.; {Mandic}, V.; {McCarthy}, K.~A.; {Mirabolfathi}, N.;
  {Moore}, D.; {Nelson}, H.; {Ogburn}, R.~W.; {Phipps}, A.; {Pyle}, M.; {Qiu},
  X.; {Ramberg}, E.; {Rau}, W.; {Reisetter}, A.; {Resch}, R.; {Saab}, T.;
  {Sadoulet}, B.; {Sander}, J.; {Schnee}, R.~W.; {Seitz}, D.~N.; {Serfass}, B.;
  {Sundqvist}, K.~M.; {Tarka}, M.; {Wikus}, P.; {Yellin}, S.; {Yoo}, J.;
  {Young}, B.~A. and {Zhang}, J.} {\em Physical Review Letters} {\bf 106},
  2011.

\bibitem{Armengaud2010}
{ E. Armengaud, C. Augier, A. Benoit, L. Berge, O. Besida, J. Blumer, A.
  Broniatowski, A. Chantelauze, M. Chapellier, G. Chardin, F. Charlieux, S.
  Collin, X. Defay, M. De Jesus, P. Di Stefano, Y. Dolgorouki, J. Domange, L.
  Dumoulin, K. Eitel, J. Gascon, G. Gerbier, M. Gros, M. Hannawald, S. Herve,
  A. Juillard, H. Kluck, V. Kozlov, R. Lemrani, P. Loaiza, A. Lubashevskiy, S.
  Marnieros, X.-F. Navick, E. Olivieri, P. Pari, B. Paul, S. Rozov, V.
  Sanglard, S. Scorza, S. Semikh, A.S. Torrento-Coello, L. Vagneron, M.-A.
  Verdier and E. Yakushev} {\em Physics Letters B} {\bf 687}, 2010.

\bibitem{Spergel2007}
D.~N. Spergel, R.~Bean, O.~Doré, M.~R. Nolta, C.~L. Bennett, J.~Dunkley,
  G.~Hinshaw, N.~Jarosik, E.~Komatsu, L.~Page, H.~V. Peiris, L.~Verde,
  M.~Halpern, R.~S. Hill, A.~Kogut, M.~Limon, S.~S. Meyer, N.~Odegard, G.~S.
  Tucker, J.~L. Weiland, E.~Wollack, and E.~L. Wright {\em The Astrophysical
  Journal Supplement Series} {\bf 170}(2), p.~377, 2007.

\bibitem{Tegmark2004}
M.~Tegmark, M.~A. Strauss, M.~R. Blanton, K.~Abazajian, S.~Dodelson,
  H.~Sandvik, X.~Wang, D.~H. Weinberg, I.~Zehavi, N.~A. Bahcall, F.~Hoyle,
  D.~Schlegel, R.~Scoccimarro, M.~S. Vogeley, A.~Berlind, T.~Budavari,
  A.~Connolly, D.~J. Eisenstein, D.~Finkbeiner, J.~A. Frieman, J.~E. Gunn,
  L.~Hui, B.~Jain, D.~Johnston, S.~Kent, H.~Lin, R.~Nakajima, R.~C. Nichol,
  J.~P. Ostriker, A.~Pope, R.~Scranton, U.~c.~v. Seljak, R.~K. Sheth,
  A.~Stebbins, A.~S. Szalay, I.~Szapudi, Y.~Xu, J.~Annis, J.~Brinkmann,
  S.~Burles, F.~J. Castander, I.~Csabai, J.~Loveday, M.~Doi, M.~Fukugita,
  B.~Gillespie, G.~Hennessy, D.~W. Hogg, i.~c.~v. Ivezi\ifmmode~\acute{c}\else
  \'{c}\fi{}, G.~R. Knapp, D.~Q. Lamb, B.~C. Lee, R.~H. Lupton, T.~A. McKay,
  P.~Kunszt, J.~A. Munn, L.~O'Connell, J.~Peoples, J.~R. Pier, M.~Richmond,
  C.~Rockosi, D.~P. Schneider, C.~Stoughton, D.~L. Tucker, D.~E. Vanden~Berk,
  B.~Yanny, and D.~G. York {\em Phys. Rev. D} {\bf 69}, p.~103501, May 2004.

\bibitem{Brink2006}
P.~L. {Brink}, B.~{Cabrera}, J.~P. {Castle}, J.~{Cooley}, L.~{Novak}, R.~W.
  {Ogburn}, M.~{Pyle}, J.~{Ruderman}, A.~{Tomada}, B.~A. {Young},
  J.~{Filippini}, P.~{Meunier}, N.~{Mirabolfathi}, B.~{Sadoulet}, D.~N.
  {Seitz}, B.~{Serfass}, K.~M. {Sundqvist}, D.~S. {Akerib}, C.~N. {Bailey},
  M.~R. {Dragowsky}, D.~R. {Grant}, R.~{Hennings-Yeomans}, and R.~W. {Schnee}
  {\em Nuclear Instruments and Methods in Physics Research A} {\bf 559},
  pp.~414--416, Apr. 2006.

\bibitem{Knoll}
G.~F. Knoll, {\em Radiation Detection and Measurement}, John Wiley and Sons,
  1989.

\bibitem{Oed1988}
{A. Oed} {\em Nuclear Instruments \& Methods in Physics Research} {\bf 263},
  p.~351, 1988.

\bibitem{Luke1994}
P.~Luke {\em Applied Physics Letters} {\bf 65}, p.~2884, 1994.

\bibitem{Agostinelli2003}
{{Geant4 Collaboration}; {Agostinelli}, S.; {Allison}, J.; {Amako}, K.;
  {Apostolakis}, J.; {Araujo}, H.; {Arce}, P.; {Asai}, M.; {Axen}, D.;
  {Banerjee}, S.; {Barrand}, G.; {Behner}, F.; {Bellagamba}, L.; {Boudreau},
  J.; {Broglia}, L.; {Brunengo}, A.; {Burkhardt}, H.; {Chauvie}, S.; {Chuma},
  J.; {Chytracek}, R.; {Cooperman}, G.; {Cosmo}, G.; {Degtyarenko}, P.;
  {dell'Acqua}, A.; {Depaola}, G.; {Dietrich}, D.; {Enami}, R.; {Feliciello},
  A.; {Ferguson}, C.; {Fesefeldt}, H.; {Folger}, G.; {Foppiano}, F.; {Forti},
  A.; {Garelli}, S.; {Giani}, S.; {Giannitrapani}, R.; {Gibin}, D.; {G{\'o}mez
  Cadenas}, J.~J.; {Gonz{\'a}lez}, I.; {Gracia Abril}, G.; {Greeniaus}, G.;
  {Greiner}, W.; {Grichine}, V.; {Grossheim}, A.; {Guatelli}, S.; {Gumplinger},
  P.; {Hamatsu}, R.; {Hashimoto}, K.; {Hasui}, H.; {Heikkinen}, A.; {Howard},
  A.; {Ivanchenko}, V.; {Johnson}, A.; {Jones}, F.~W.; {Kallenbach}, J.;
  {Kanaya}, N.; {Kawabata}, M.; {Kawabata}, Y.; {Kawaguti}, M.; {Kelner}, S.;
  {Kent}, P.; {Kimura}, A.; {Kodama}, T.; {Kokoulin}, R.; {Kossov}, M.;
  {Kurashige}, H.; {Lamanna}, E.; {Lamp{\'e}n}, T.; {Lara}, V.; {Lefebure}, V.;
  {Lei}, F.; {Liendl}, M.; {Lockman}, W.; {Longo}, F.; {Magni}, S.; {Maire},
  M.; {Medernach}, E.; {Minamimoto}, K.; {Mora de Freitas}, P.; {Morita}, Y.;
  {Murakami}, K.; {Nagamatu}, M.; {Nartallo}, R.; {Nieminen}, P.; {Nishimura},
  T.; {Ohtsubo}, K.; {Okamura}, M.; {O'Neale}, S.; {Oohata}, Y.; {Paech}, K.;
  {Perl}, J.; {Pfeiffer}, A.; {Pia}, M.~G.; {Ranjard}, F.; {Rybin}, A.;
  {Sadilov}, S.; {di Salvo}, E.; {Santin}, G.; {Sasaki}, T.; {Savvas}, N.;
  {Sawada}, Y.; {Scherer}, S.; {Sei}, S.; {Sirotenko}, V.; {Smith}, D.;
  {Starkov}, N.; {Stoecker}, H.; {Sulkimo}, J.; {Takahata}, M.; {Tanaka}, S.;
  {Tcherniaev}, E.; {Safai Tehrani}, E.; {Tropeano}, M.; {Truscott}, P.; {Uno},
  H.; {Urban}, L.; {Urban}, P.; {Verderi}, M.; {Walkden}, A.; {Wander}, W.;
  {Weber}, H.; {Wellisch}, J.~P.; {Wenaus}, T.; {Williams}, D.~C.; {Wright},
  D.; {Yamada}, T.; {Yoshida}, H. and {Zschiesche}, D.} {\em Nuclear
  Instruments and Methods in Physics Research A} {\bf 506}, 2003.

\bibitem{RPP}
{K. Nakamura et al. (Particle Data Group)} {\em Journal of Physics G} {\bf 37},
  2010.

\bibitem{Cabrera1993}
B.~{Cabrera}, B.~L. {Dougherty}, A.~T. {Lee}, M.~J. {Penn}, J.~G. {Pronko}, and
  B.~A. {Young} {\em Journal of Low Temperature Physics} {\bf 93}({3-4}),
  pp.~365--375, 1993.

\bibitem{Fano1947}
U.~{Fano} {\em Physical Review} {\bf 72}, pp.~26--29, July 1947.

\bibitem{Rutherford1911}
E.~Rutherford {\em Philos. Mag.} {\bf 6}, p.~21, 1911.

\bibitem{Bonderup1978}
E.~Bonderup {\em Lectures notes, Aarhus} , 1978.

\bibitem{Lindhard1954}
{J. Lindhard} {\em Matematisk-fysiske Meddelelser udgivet af Det Kongelige
  Danske Videnskabernes Selskab} {\bf 28}(8), 1954.

\bibitem{Lindhard1963}
{J. Lindhard, V. Nielsen, M. Scharff and P.V. Thomsen} {\em Matematisk-fysiske
  Meddelelser udgivet af Det Kongelige Danske Videnskabernes Selskab} {\bf
  33}(10), 1963.

\bibitem{Lindhard1963_2}
{J. Lindhard, M. Scharff and H.E. Schiott} {\em Matematisk-fysiske Meddelelser
  udgivet af Det Kongelige Danske Videnskabernes Selskab} {\bf 33}(14), 1963.

\bibitem{Jones1975}
{K.W. Jones and H.W. Kraner} {\em Physical Review A} {\bf 11}(4),
  pp.~1347--1353, 1975.

\bibitem{Lewin1996}
J.~Lewin and P.~Smith {\em Astroparticle Physics} {\bf 6}, pp.~87--112, 1996.

\bibitem{Kittel}
C.~Kittel and H.~Kroemer, {\em Thermal Physics}, W.H. Freeman and Company,
  1980.

\bibitem{Ashcroft}
N.~W. Ashcroft and N.~D. Mermin, {\em Solid State Physics}, Harcourt Collee
  Publishers, 1976.

\bibitem{Wolfe}
J.~P. Wolfe, {\em Imaging Phonons: Acoustic Wave Propagation in Solids},
  Cambridge University Press, 1998.

\bibitem{Nye}
J.~F. Nye, {\em Physical Properties of Crystals}, Oxford University Press,
  1957.

\bibitem{Northrop1980}
G.~A. {Northrop} and J.~P. {Wolfe} {\em Physical Review B} {\bf 22},
  pp.~6196--6212, Dec. 1980.

\bibitem{Tamura1991}
S.~{Tamura}, J.~A. {Shields}, and J.~P. {Wolfe} {\em Physical Review B} {\bf
  44}, pp.~3001--3011, Aug. 1991.

\bibitem{Maris1993}
H.~J. {Maris} and S.~{Tamura} {\em Physical Review B} {\bf 47}, pp.~727--739,
  Jan. 1993.

\bibitem{Tamura1993LT}
S.~{Tamura} {\em Journal of Low Temperature Physics} {\bf 93}(3), pp.~433--438,
  1993.

\bibitem{Maris1990}
J.~{Maris} {\em Physical Review B} {\bf 41}, pp.~9736--9743, May 1990.

\bibitem{Msall1993}
M.~{Msall}, S.~{Tamura}, S.~{Esipov}, and J.~{Wolfe} {\em Physical Review
  Letters} {\bf 70}, pp.~3463--3466, May 1993.

\bibitem{Tamura1993}
S.~{Tamura} {\em Physical Review B} {\bf 48}, pp.~13502--13507, Nov. 1993.

\bibitem{NumericalRecipes}
W.~H. Press, S.~A. Teukolsky, W.~T. Vetterling, and B.~P. Flannery, {\em
  Numerical Recipes in C: The Art of Scientific Computing, Second Edition},
  Cambridge University Press, 1992.

\bibitem{Tamura1985TA}
{Shin-ichiro Tamura} {\em Physical Review B} {\bf 31}(4), 1985.

\bibitem{Tamura1985}
{Shin-ichiro Tamura} {\em Physical Review B} {\bf 31}(4), 1985.

\bibitem{CabreraTamura}
{Private correspondence between Blas Cabrera and Shin-ichiro Tamura}.

\bibitem{Leman2011}
S.~Leman, B.~Cabrera, K.~McCarthy, M.~Pyle, R.~Resch, B.~Sadoulet,
  K.~Sundqvist, P.~Brink, M.~Cherry, {Do Couto E Silva, E.},
  E.~Figueroa-Feliciano, N.~Mirabolfathi, B.~Serfass, and A.~Tomada {\em
  Chinese Journal of Physics (PHONONS 2010 conference proceedings)} {\bf 49},
  p.~349, 2011.

\bibitem{Leman2011_4}
{S.W. Leman}, K.~McCarthy, P.~Brink, B.~Cabrera, M.~Cherry, {E. Do Couto E
  Silva}, E.~Figueroa-Feliciano, P.~Kim, N.~Mirabolfathi, M.~Pyle, R.~Resch,
  B.~Sadoulet, B.~Serfass, K.~Sundqvist, , A.~Tomada, , and B.~Young {\em
  Journal of Applied Physics} {\bf 110}, p.~094515, 2011.

\bibitem{Pehl1968}
R.~H. {Pehl}, F.~S. {Goulding}, D.~A. {Landis}, and M.~{Lenzling} {\em Nuclear
  Instruments and Methods} {\bf 59}(1), pp.~45--55, 1968.

\bibitem{Kaplan1976}
{S.B. Kaplan, C.C. Chi, D.N. Langenber, J.J. Chang, S. Jafarey and D.J.
  Scalapino} {\em Physical Review B} {\bf 14}(11), p.~4854, 1976.

\bibitem{Kurakado1982}
M.~Kurakado {\em Nuclear Instruments and Methods} {\bf 196}, pp.~275--277,
  1982.

\bibitem{BrinkThesis}
P.~L. Brink, {\em Non-Equilibrium Superconductivity induced by X-ray Photons}.
\newblock PhD thesis, Oxford Universiy, 1995.

\bibitem{Sasaki1958}
{W. Sasaki, M. Shibuya, K. Mizuguchi } {\em Journal of the Physical Society of
  Japan} {\bf 13}, p.~456, 1958.

\bibitem{Jacoboni1983}
C.~Jacoboni and L.~Reggiani {\em Rev. Mod. Phys.} {\bf 55}, pp.~645--705, Jul
  1983.

\bibitem{AubryFortuna2010}
{V. Aubry-Fortuna and P. Dollfus} {\em Journal of Applied Physics} {\bf 108},
  2010.

\bibitem{Neganov1985}
B.~Neganov and V.~Trofimov {\em USSR Patent No 1037771, Otkrytia i
  izobreteniya} {\bf 146}, p.~215, 1985.

\bibitem{Luke1988}
P.~Luke {\em Journal of Applied Physics} {\bf 64}, p.~6858, 1988.

\bibitem{Herring1956}
C.~Herring and E.~Vogt {\em Physical Review} {\bf 101}(3), p.~944, 1956.

\bibitem{Verlet1967}
{Loup Verlet} {\em Physical Review} {\bf 159}(1), p.~98, 1967.

\bibitem{Swope1982}
{William C. Swope, Hans C. Andersen, Peter H. Berens and Kent R. Wilson} {\em
  Journal of Chemical Physics} {\bf 76}, p.~637, 1982.

\bibitem{Sundqvist2009}
{K.M. Sundqvist, A. T. J. Phipps, C. N. Bailey, P. L. Brink, B. Cabrera, M.
  Daal, A. C. Y. Lam, N. Mirabolfathi, L. Novak, M. Pyle, B. Sadoulet, B.
  Serfass, D. Seitz, A. Tomada and J. J. Yen} {\em AIP Conference Series} {\bf
  1185}, pp.~128--131, 2009.

\bibitem{cgal}
www.cgal.org.

\bibitem{KIrwin1995}
K.~D. Irwin {\em Applied Physics Letters} {\bf 66}(15), 1995.

\bibitem{Irwin2005}
{Irwin, K.D. and Hilton, G.C.}, ``{Transition-Edge Sensors},'' in {\em
  {Cryogenic Particle Detection}},  {Enss, Christian}, ed., {\em {Topics in
  Applied Physics}} {\bf {99}}, pp.~{81--97}, {Springer Berlin / Heidelberg},
  {2005}.

\bibitem{Cabrera2000}
B.~Cabrera {\em Nuclear Instruments \& Methods in Physics Research, Section A}
  {\bf 444}(1-2), 2000.

\bibitem{LemanThesis}
S.~W. Leman, {\em Development of Phonon-Mediate Transition-Edge-Sensor X-ray
  Detectors for use in Astronomy}.
\newblock PhD thesis, Stanford Universiy, 2001.

\bibitem{Anderson2011}
{A.J. Anderson, S.W. Leman, M. Pyle, E. Figueroa-Feliciano, K. McCarthy and T.
  Doughty for the SuperCDMS Collaboration} {\em Journal of Low Temperature
  Physics} {\bf 167}, pp.~135--140, 2012.

\bibitem{Anderson2011_2}
{A.J. Anderson, S.W. Leman, T. Doughty, E. Figueroa-Feliciano, K.A. McCarthy,
  M. Pyle and B.A. Young} {\em Nuclear Instruments and Methods in Physics
  Research A} , submitted 2011.

\bibitem{Tinkham}
M.~Tinkham, {\em Introdution to Superconductivity}, Krieger Publishing Company,
  1975.
\newblock see page 117.

\bibitem{Irwin1998}
K.~Irwin, G.C.Hilton, D.A.Wollman, and J.~M. Martinis {\em Journal of Applied
  Physics} {\bf 83}(8), 1998.

\bibitem{Saab2000}
{T. Saab, R.M. Clarke, B. Cabrera, R.A. Abusaidi and R. Gaitskell } {\em
  Nuclear Instruments \& Methods in Physics Research A} {\bf 444}, p.~300,
  2000.

\bibitem{Pyle2004}
{M. Pyle, P.L. Brink, B. Cabrera, J.P. Castle, P. Colling, C.L. Chang, J.
  Cooley, T. Lipus, R.W. Ogburn and B.A. Young} {\em Nuclear Instruments \&
  Methods in Physics Research A} {\bf 559}, p.~405, 2006.

\bibitem{Courant1928}
R.~{Courant}, K.~{Friedrichs}, and H.~{Lewy} {\em Mathematische Annalen} {\bf
  100}, pp.~32--74, 1928.

\bibitem{Burden}
R.~L. Burden and J.~D. Faires, {\em Numerical Analysis}, Brooks / Cole
  Publishing Company, 1997.

\bibitem{Leman2011_2}
{S.W. Leman}, S.~Hertel, P.~Kim, B.~Cabrera, {E. Do Couto E Silva},
  E.~Figueroa-Feliciano, K.~McCarthy, R.~Resch, B.~Sadoulet, K.~Sundqvist, and
  on~behalf of~the Cryogenic Dark Matter Search~collaboration {\em Journal of
  Low Temperature Physics} {\bf 167}, pp.~1106--1111, 2012.

\bibitem{Leman2011_3}
{S.W. Leman}, D.~Brandt, P.~Brink, B.~Cabrera, H.~Chagani, M.~Cherry,
  P.~Cushman, {E. Do Couto E Silva}, T.~Doughty, E.~Figueroa-Feliciano,
  V.~Mandic, K.~McCarthy, N.~Mirabolfathi, M.~Pyle, A.~Reisetter, R.~Resch,
  B.~Sadoulet, B.~Serfass, K.~Sundqvist, A.~Tomada, B.~Young, J.~Zhang, and
  on~behalf of~the Cryogenic Dark Matter Search~collaboration {\em Journal of
  Low Temperature Physics} {\bf 167}, pp.~1099--1105, 2012.

\bibitem{McCarthy2011}
{K.A.~McCarthy, S. W.~Leman, A.~Anderson, D.~Brandt, P.L.~Brink, B.~Cabrera,
  M.~Cherry, E.~Do~Couto~E~Silva, P.~Cushman, T.~Doughty,
  E.~Figueroa-Feliciano, P. Kim, N.~Mirabolfathi, L.~Novak, R.~Partridge,
  M.~Pyle, A. Reisetter, R.~Resch, B.~Sadoulet, B.~Serfass, K.M.~Sundqvist and
  A.~Tomada} {\em Journal of Low Temperature Physics} {\bf 167},
  pp.~1160--1166, 2012.

\bibitem{Broniatowski2004}
A.~Broniatowski {\em Nuclear Instruments and Methods in Physics Research A}
  {\bf 520}, pp.~178--181, 2004.

\bibitem{Broniatowski2011}
{A. Broniatowski (EDELWEISS collaboration)} {\em Journal of Low Temperature
  Physics} {\bf 167}, pp.~1069--1074, 2012.

\bibitem{Brandt2011}
D.~Brandt, M.~Asai, P.~Brink, B.~Cabrera, E.~do~Couto~e Silva, M.~Kelsey,
  S.~Leman, K.~McCarthy, R.~Resch, and D.~Wright {\em Journal of Low
  Temperature Physics} {\bf 167}, pp.~485--490, 2012.

\end{thebibliography}


\end{document}